\RequirePackage{lineno}
\documentclass[showpacs,aps,twocolumn]{revtex4-1}

\usepackage{bm}
\usepackage{amsmath}
\usepackage{graphicx}
\usepackage{subfigure}
\usepackage[usenames,dvipsnames]{color}
\definecolor{darkblue}{RGB}{0,0,196}
\usepackage[colorlinks=true,linkcolor=darkblue,citecolor=darkblue,urlcolor=darkblue]{hyperref}

\usepackage{setspace}
\usepackage{hyperref}
\usepackage{footmisc}
\usepackage[makeroom]{cancel}
\usepackage{comment}
\usepackage{lineno}
\usepackage{ulem}

\def\be{\begin{equation}}
\def\ee{\end{equation}}
\def\ba{\begin{eqnarray}}
\def\ea{\end{eqnarray}}

\usepackage{graphicx}
\usepackage{amsmath,bbm}
\usepackage{amssymb}
\usepackage{yfonts}

\begin{document}
\title{Elliptic flow of hadrons via quark coalescence mechanism using Boltzmann transport equation for Pb+Pb collision at $\sqrt{s_{NN}}$=2.76 TeV}
\author{Mohammed Younus}
\author{Sushanta Tripathy}
\author{Swatantra Kumar Tiwari}
\author{Raghunath~Sahoo\footnote{Corresponding author: $Raghunath.Sahoo@cern.ch$}}
\affiliation{Discipline of Physics, School of Basic Sciences, Indian Institute of Technology Indore, Simrol, Indore 453552, India}

\begin{abstract}
\noindent
{\bf Abstract:} Elliptic flow of hadrons observed at relativistic heavy-ion collision experiments at Relativistic Heavy-Ion Collider (RHIC) and Large Hadron Collider (LHC), provides us an important signature of  possible de-confinement transition from hadronic phase to partonic phase. However, hadronization processes of de-confined partons back into final hadrons are found to play a vital role in the observed hadronic flow. In the present work, we use coalescence mechanism also known as Recombination (ReCo) to combine quarks into hadrons. To get there, we have used Boltzmann transport equation in relaxation time approximation to transport the quarks into equilibration and finally to freeze-out surface, before coalescence takes place. A Boltzmann-Gibbs Blast Wave (BGBW) function is taken as an equilibrium function to get the final distribution and a power-like function to describe the initial distributions of partons produced in heavy-ion collisions. In the present work, we try to estimate the elliptic flow of identified hadrons such as $\pi$, $K$, $p$ etc., produced in Pb+Pb collisions at $\sqrt{s_{\rm NN}}$ = 2.76 TeV at the LHC for different centralities.  The elliptic flow ($v_2$) of identified hadrons 
seems to be described quite well in the available $p_{\rm T}$ range. After the evolution of quarks until freeze-out time, has been calculated using BTE-RTA, the approach used in this paper consists of combining two or more quarks to explain the produced hadrons at intermediate momenta regions. The formalism is found to describe elliptic flow of hadrons produced in Pb+Pb collisions to a large extent.


\end{abstract}
\date{\today}
\maketitle 
\section{Introduction}
\label{intro}
Study of collective phenomena, among many contemporary  signatures of quark-gluon-plasma (QGP), continues to remain in the forefront of scientific investigations~\cite{voloshin,LeFevre:2016vpp}. While, on the experimental front~\cite{v2_star1,v2_star2,v2_star3,v2_star4,v2_phenix1,
v2_phenix2,v2_ALICE1} analysis of final hadrons' data from RHIC and LHC experiments has enabled us to look back in time and reconstruct the flow phenomena, the phenomenological models using theoretical and numerical techniques have been able to simulate the events, starting from the point of collision of heavy ions, until the freeze-out. The theoretical results have been successful in explaining experimental data to a large extent. With the advent of new techniques, the time is however ripe to be able to resolve differences in theories and experiments, and precisely determine various observables of QGP.

Earlier, attempts were made through extensive theoretical modelling and analysis of data, to reconstruct azimuthal anisotropy, (also known as elliptic flow) $v_2$, of hadrons in transverse momentum plane~\cite{Biro:2015iua,Nahrgang:2014vza,Haque:2017qhe,Tripathy:2017nmo}. However, it is believed that azimuthal anisotropy would develop at the early phase of heavy ion collision, when bulk of the de-confined quarks and gluons from non-central collision between two heavy ions, goes into local thermalization or quark gluon plasma (QGP) state. The geometrical asymmetry of the spatial overlap zone is transformed into momentum anisotropy of the produced particles. With the onset of the local themalization of the bulk partonic matter, the azimuthal anisotropy (mathematically, the second coefficient of the Fourier expansion of particle transverse momentum spectrum), is exhibited strongly in the collective behaviour of the quark-gluon-plasma. This information on initial anisotropy is carried till freeze-out, and finally reflected in the hadron spectra. However, the rapid expansion of the medium towards isotropization may smear this information to some extent. But on the other hand, hadronic medium effects may add to the partonic flow until kinetic freeze-out sets in. Thus it is necessary to develop robust calculation as to able to discern factors emanating from various phases of heavy ion collision, which may affect particles' flow. While phenomenologies of hadronic matter try to reconstruct the $v_2$ from the final hadronic spectra, initial anisotropy in the partons' configuration space on the other hand affects the formation of flow and is calculated using phenomenological models such as Glauber mechanism along with perturbative QCD based calculations. However, the two phases of initial anisotropy and hadrons' interaction remain separated by QGP phase. As mentioned  earlier, QGP phase contributes to the evolution of particle flow to a great extent. Hence, It is up to the transport models which may properly bring in the QGP effects and bridge the initial anisotropy and effects of hadronic phase in the observed $v_2$~\cite{Sun:2015pta}. The transport models help us in studying collision centrality dependency of QGP properties. They not only shed light on properties of hot and dense matter viz. average temperature and momentum reached by equilibrated system and their dependency on collision centrality, but also provide us with vast information on transport properties such as radial flow coefficients, momentum broadening, drag and diffusion coefficients, electrical and thermal conductivity of QGP matter etc~\cite{Teaney:2000cw,Mazumder:2013oaa,Sarkar:2017fni,Guo:2017mkf,Sahoo:2018dxn}. 

The available transport calculations are based on either hydrodynamical equations, Langevin equation, or Boltzmann transport equation. In this paper, we have used Boltzmann transport equation (BTE) in relaxation time approximation (RTA). BTE-RTA would transport the entire parton distribution to equilibration and then to freeze-out surface where-after kinetic or chemical interaction among particles ceases completely. Using BTE-RTA approach in the present work, we have attempted to study parameters associated with particle production such as,  radial flow and relaxation time of the final state particles  etc. Neglecting the effects of the hadron medium on the particles' flow as approximation, the current work focuses on the interplay of the various parameters and mechanisms on the partonic states. We will discuss our approach in detail in subsequent sections. We have also assumed that the final quarks would hadronize into mesons and baryons using partonic coalescence mechanism at the hadronization hypersurface. We will discuss this formalism in one of the following sections. Finally, elliptic flow, $v_2$, for various hadrons is calculated and presented in the results and discussion section. We have also presented figures on our study of the parameters and their inter-dependencies. We have then concluding section of our paper, which is followed by bibliography.

Let us now discuss BTE-RTA formalism briefly.

\noindent
\section{Relaxation time approximation (RTA) of Boltzmann transport equation (BTE)}
\label{bte}
As mentioned in the introductory section, the evolution of quarks within the medium towards the freeze-out surface have major effects on the observed final particle spectra. The transport calculations such as hydrodynamics, BTE etc. are commonly used as the evolution mechanisms and provide description of hadron spectra in both qualitative and quantitative manner~\cite{Baier:2006um,Gavin:1985ph,Geiger:1991nj,Srivastava:1997qg,Bass:2004vh,Zhang:1999rs,Younus:2013rja}. We know that various dynamical features ranging from multi-parton interaction, in-medium energy loss, thermal, and chemical equilibrations, to dynamics at freeze-out surfaces contribute extensively to the particle flow and can be studied using BTE. We also know that partons evolving through space and time undergo several collisions and thermalize. Furthermore, they continue to evolve and expand until freeze-out even after hadronization. Any of these features can be studied using BTE.

The BTE in general can be written as:
\begin{eqnarray}
\label{eq2}
 \frac{df(x,p,t)}{dt}=\frac{\partial f}{\partial t}+\vec{v}.\nabla_x
f+\vec{F}.\nabla_p
f=C[f],
\end{eqnarray}
where $f(x,p,t)$ is the distribution of particles which depends on position, momentum and time. $\vec{v}$ is the velocity and $\vec{F}$ is the external force. $\nabla_x$ and $\nabla_p$ are the partial derivatives with respect to position and momentum, respectively. $C[f]$ is the collision term which depicts the interaction of the particles with the medium or among themselves. Earlier, BTE has also been used in RTA to study the time evolution of temperature fluctuation in a non-equilibrated system \cite{Bhattacharyya:2015nwa} and also for studying the $R_{AA}$ and $v_2$ of various light and heavy flavours at RHIC and LHC energies~\cite{Tripathy:2016hlg,Tripathy:2017kwb}. 

We have considered the evolution of particle momentum distribution with time. We have taken $\nabla_x f=0$ assuming particle distribution to be homogeneous in space and the configuration space distribution or spatial distribution has been parametrized accordingly. There aren't any external forces acting on the system ($\vec{F}=$0). Hence, the second and third terms of eq.~\eqref{eq2} become zero and it reduces to,
\ba
 \label{eq3}
  \frac{df(x,p,t)}{dt}=\frac{\partial f}{\partial t}=C[f].
\ea

The full kernel of the collision term $\displaystyle{C[f]}$ contains microscopic interaction cross-sections of particles. For any transport models such as AMPT, UrQMD etc. containing microscopic Boltzmann equation, full interaction kernel along with space and time evolution of the system becomes important. In our calculation owing to assumed homogeneous spatial distribution, the spatial variables have been parametrized. 

In BTE-RTA \cite{Florkowski:2016qig} which is an effective model, the collision term is however expressed as:
\ba
\label{eq4}
 C[f] =-\frac{f-f_{eq}}{\tau},
 \label{colltermrta}
\ea
where $f_{eq}$ is Boltzmann local equilibrium distribution characterized by a freeze-out temperature $T$. $\tau$ is the relaxation time, the time taken by a non-equilibrium system to reach equilibrium. Using eq.~\eqref{colltermrta}, eq.~\eqref{eq3} becomes 
\ba
 \label{eq5}
  \frac{\partial f^q}{\partial t}=-\frac{f^q-f^q_{eq}}{\tau}.
\ea
Solving the above equation with the initial conditions {\it i.e.} at $t=0, f=f_{i}$ and at $t=t_f, f=f_{f}$, in general, we get final distribution for any quark flavour as:
\ba
 \label{eq6}
 f^q_{f}=f^q_{eq}+(f^q_{i}-f^q_{eq})e^{-\frac{t_f}{\tau}},
\ea
where $t_f$ is the freeze-out time parameter. The initial distribution $\displaystyle{f_i}$ at $\displaystyle{t\,=\,0}$ is taken as power-like distribution. We will come back to this later. BTE-RTA computes to give the final distribution $\displaystyle{f_f}$ as function of parameter $\displaystyle{t_f/\tau}$. If system is given enough time or $\displaystyle{t_f}$ is large compared to $\displaystyle{\tau}$, $\displaystyle{f_f}$ might converge to $\displaystyle{f_{eq}}$.

We use eq.~\eqref{eq6} in the definition of the elliptic flow ($v_2$) at mid-rapidity, which is expressed as,
\ba
\label{eq7}
v_2^q(p_T)=\frac{\int{f^q_{f} \times \cos(2\phi)\,d\phi}}{\int{f^q_{f}\,d\phi}}.
\ea
Eq.~\eqref{eq7} gives azimuthal anisotropy after incorporating RTA in BTE. Boltzmann-Gibbs Blast Wave (BGBW) function has been taken as equilibrium distribution function, $f_{eq}$, as:
\ba
\label{bgbw1}
f^q_{eq}(p_T)&=&C.exp(-\frac{p^\mu u_\mu}{T})\nonumber\\
\frac{dN^{eq}_{q/\bar{q}}}{p_Tdp_Tdy}&=&\int d^3\sigma_\mu p^\mu f^{q/\bar{q}}_{eq}(p_T)\,,
\ea
where the particle four-momentum is, $p^\mu = (m_T\cosh y, p_T\cos\phi, p_T\sin\phi, m_T\sinh y)$, the four-velocity denoting flow velocities in space-time is given by, $u^\mu = \cosh\rho(\cosh\eta, \tanh\rho \cos\phi_r, \tanh\rho \sin \phi_r, \sinh \eta)$, while the kinetic freeze-out surface is given by $d^3\sigma_\mu = (\cosh\eta, 0, 0, -\sinh\eta)\tau rdrd\eta d\phi_r$. Here, $\eta$ is the space-time rapidity. Assuming boost-invariant scenario where we have taken Bjorken correlation in rapidity, $i.e.$ $y=\eta$~\cite{Bjorken:1982qr} along longitudinal or beam axis. Thus, eq.~\eqref{bgbw1} can be expressed as
\begin{widetext}  
\ba
\label{feq}
\centering
\frac{dN^{eq}_{q/\bar{q}}}{p_Tdp_Tdy}= \frac{1}{2\pi}.D \int_0^{R_{0}} r\;dr\;\int_0^{\infty} {\cosh}y\;{\exp}\Big(-\frac{m_T\;{\cosh}y\;{\cosh}\rho}{T}\Big)dy\;\int_0^{2\pi} exp\Big(\frac{p_T\;{\sinh}\rho\;{\cos}\phi}{T}\Big)d\phi,
\ea
\end{widetext}
where $D = \displaystyle \frac{g_q\,t_f\,m_T}{2\pi^2}$. Here $g$ is the quark degeneracy factor, $t_f$ is the particle emission time, and $m_{\rm T}=\sqrt{p_T^2+m^2_q}$ is the transverse mass.

$\rho$ in the integrand is a transverse rapidity variable which is given by $\rho=tanh^{-1}\beta_r+\rho_a(b) \cos(2\phi)$, with $\rho_a$ as a function of impact parameter, $b$ and gives the anisotropy dependence in the flow. $\beta_r=\displaystyle\beta_s\;\xi^n$ \cite{Huovinen:2001cy,Schnedermann:1993ws,BraunMunzinger:1994xr, Tang:2011xq} is the radial flow, where $\beta_s$ is the maximum surface velocity and $\xi=\displaystyle r/R_0$, with $r$ as the radial distance from the center of the fireball. In the blast-wave model, the particles closer to the center of the fireball move slower than the ones at the edges. The average of the transverse velocity can be evaluated as \cite{Adcox:2003nr}, 
\ba
<\beta_r> =\frac{\int \beta_s\,\xi^{n+1}\;d\xi}{\int \xi\;d\xi}=\Big(\frac{2}{2+n}\Big)\beta_s.
\ea
While the anisotropic parameter, $\rho_a$ is written as,
\ba
\rho_a(b)=\left[\frac{\sqrt{1-\zeta^2}-(1-\zeta)}{\sqrt{1-\zeta^2}+(1-\zeta)}\right]\,,\,\zeta=\frac{b}{2R_a}\,.
\ea
In our calculation, we use a linear velocity profile, ($n=1$), $R_0$ is the maximum radius of the expanding source at freeze-out ($0<\xi<1$), and $R_A$ is the radius of colliding nucleus. $b$ is the impact parameter to include the centrality dependence of anisotropy. In this paper, we have parametrized the initial distribution given by particle production using perturbative QCD leading order (pQCD LO) calculations for $p+p$ collision,

\begin{eqnarray}
\label{pqcd1}
\frac{d\sigma_{pp\rightarrow q\bar{q}}}{d^2p_Tdy_1dy_2}&=&2x_1x_2 \sum_{1,2}\bigg[f_p(x_1,Q^2).f_{p}(x_2,Q^2).\frac{d\hat{\sigma}_{12\rightarrow q\bar{q}}}{d\hat{t}} \nonumber\\
 &+&(1\leftrightarrow 2)\bigg]\times\frac{1}{\left(1+\delta_{12}\right)}\,.
\end{eqnarray}
Here, $x_1$ and $x_2$ are momentum fractions carried by interacting partons from their respective colliding protons and are given by, 
\ba
x_1&=&\frac{2m_T}{\sqrt{s}}\left(\exp{(-y_1)}+\exp{(-y_2)}\right)\,,\nonumber\\
x_2&=&\frac{2m_T}{\sqrt{s}}\left(\exp{(-y_1)}-\exp{(-y_2)}\right)\,.
\ea
A $p_T$ cut of 2 GeV/c is taken for the jet production following other event generators like - PYTHIA, HIJING etc. \cite{pythia,hijing}.
The parton density functions, $f_i(x,Q^2)$, are taken to be CTEQ5M~\cite{Lai:1996mg}. The partonic differential scattering cross-sections, $\displaystyle\frac{d\hat{\sigma}}{d\hat{t}}$ is calculated from the LO processes, $gg\rightarrow q\bar{q}$ and $q\bar{q}\rightarrow q\bar{q}$. To incorporate NLO processes, we have taken a factor, $'K'$, of value 2.5, and finally nuclear overlap function, $T_{AA}(b)$ and EKS98 parametrization for shadowing effects are taken into account to convert particle production cross-section from $p+p$ collision into $A+A$, particle spectra. Eq.~\ref{pqcd1} is parametrized using a function with a power-like structure (Juttner distr.) and we fixed the parameters of the pre-equilibrated partons shown in Table~\ref{table}. However, it is worthwhile to mention that other types of functions can be utilized to obtain the initial quark distributions.

\ba
\label{eqfi}
\frac{dN_{pp\rightarrow q\bar{q}}}{d^2p_Tdy_1dy_2}&=&T_{AA}(b).\frac{d\sigma_{pp\rightarrow q\bar{q}}}{d^2p_Tdy_1dy_2}\nonumber\\
&=&T_{AA}(b).K.C.\left[1+{\frac{m_T}{B}}\right]^{-\alpha}\nonumber\\
f_i^{q/\bar{q}}(p_T)&=&\frac{1}{t_f.\pi.\,R^2_A.m_T.\cosh(y-\eta)}\frac{dN^i_{q/\bar{q}}}{d^2p_Tdy}\nonumber\\
\ea 

Here too, we have assumed Bjorken correlation in rapidity, $i.e.$ $y=\eta$. Using Glauber model, $T_{AA}(b)$ is calculated to be $260.50\,fm^{-2}$ for 0-5\% centrality, and $13.1\,fm^{-2}$ for 50-60\% centrality of the colliding nuclei at $\sqrt{s_{NN}}$ = 2.76 TeV. Using eqs.~\eqref{feq} and \eqref{eqfi}, the final distribution can be expressed as in eq.~\eqref{eq6}. This gives the final $p_T$ distribution for quarks. The quark masses for the initial distributions is taken to be, $\displaystyle m_u\,=\,2.3$ MeV, $\displaystyle m_d\,=\,4.5$ MeV, $\displaystyle m_s\,=\,95$ MeV, and $\displaystyle m_c\,=\,1.25$ GeV. After the transport quark coalescence formalism has been used to combine the quarks into hadrons. This will be discussed next.

\noindent
\section{Quark coalescence}
\label{reco}
The quark coalescence model (ReCo) is used to recombine quarks into hadrons and is found to be one of the prominent hadronization mechanisms beside parton fragmentation~\cite{Molnar:2003ff,Lin:2003jy}. 
In Refs. \cite{Fries:2003vb,Fries:2003kq}, the authors have used a two component behaviour of hadronic spectra.  For low-$p_{T}$ ($p_{T} < $ 5 GeV/c), they have used the recombination mechanism for thermalized partons, while for $p_{T} > $ 5 GeV/c a power law-like distribution with fragmentation has been used. In our work, we have concentrated our transport approach to low and intermediate $p_T$ ($<$ 5 GeV/c), where instead of adopting thermalized distribution directly, we have allowed jet distribution of partons to relax or thermalize and then proceed to hadronization. The idea was use of BTE-RTA equation to study the applicability of interpolation of jet distribution with Blast wave equation at the intermediate p$_T$ region. The coalescence or recombination of partons into hadrons has been able to explain experimentally observed hadron spectra in the intermediate and perhaps at the low momentum regions, while parton fragmentation processes are aptly suitable in explaining hadrons with high momenta. And thus ReCo mechanism has been used for final transported quark distributions in the present work. ReCo mechanism also highlights the major contribution of partonic degrees of freedom in observed hadron flow. 
The process such as $\textstyle{g\,g\,\rightarrow \,g\,g}$ has been neglected as gluons contribution are mostly at low $\displaystyle{p_T\,<\,}$ 1 GeV/c. While high $\displaystyle{p_T}$ gluons contribute to hadrons via fragmentation mechanism. In the intermediate momentum region, constituent quark counting becomes important for recombination process. At the hadronization surface only constituent quarks behave as effective degrees of freedom with mass. However it must be noted that for net entropy and energy density calculation, gluon contribution is most vital~\cite{Fries:2003vb,Fries:2003kq,Kaczmarek:1999mm,Peshier:2002ww,Casalderrey:2003cf}. In the present work we haven't included gluon contribution to the hadron production which is one of the main difference from the earlier works and it is most visible at the low momenta where gluon contribution is important. Unlike earlier works we haven't included fragmentation mechanism for high p$_T$ particles and only focussed our observations to the intermediate p$_T$. Another difference in current work is absence of hadron decay mechanism which is important at low momentum region. However, these are out of the scope of the current work and could constitute a better prospective for future research.

The coalescence model can be applied to the quarks at the hadronization surface when two (three) quarks recombine to form mesons (baryons)~\cite{Greco:2003mm, He:2017tla,Sun:2017ooe}. The model can be further utilized in describing observed spectra of light nuclei such as deuteron which contains a neutron and a proton~\cite{Yin:2017qhg}.

The coalescence model combines two or more quark distributions using convoluting functions also known as Wigner functions. The basic equation showing the number of mesons from two combining quarks can be broadly written as,

\ba
\label{reco1}
N_M=&g_M&\int{m_{T1}\cosh(y_1-\eta_1)d^3r_1}\nonumber\\
&\times&m_{T2}\cosh(y_2-\eta_2)d^3r_2\,d\vec{p}_{T1}dy_1\,d\vec{p}_{T2}dy_2\nonumber\\
&\times&f_q(r_1;p_1)\,f_{\bar{q}}(r_2;p_2)\,W_M(r_1,r_2;p_1,p_2)\,.\nonumber\\
\ea
 where, $\vec{r_1},\vec{r_2}$ and $\vec{p}_{T1},\vec{p}_{T2}$ are the spatial and transverse momentum coordinates of the combining quarks and anti-quarks, $f_{q/\bar{q}}$ are the quark distribution functions. $W_M(r_1,r_2;p_1,p_2)$ is the Wigner function convoluting two partonic distributions. $g_M$ in the front of eq.~\ref{reco1}, is meson degeneracy factor. We have also assumed Bjorken correlation in rapidities, $\displaystyle y_1=\eta_1$ and $\displaystyle y_2=\eta_2$, throughout. We also assumed $|y_1|\,=\,|y_2|$ $\leq$ 0.5. This ensures a close phase space for quarks in both momentum and configuration spaces.
 
We have assumed the delta functions correlation, $\delta^3{(\vec{p}-\vec{p}_{1}-\vec{p}_{2})}$, and $\delta^3{(\vec{2R}-\vec{r_1}-\vec{r_2})}$. We have defined the partons in the spatial and momentum coordinates in the C.M. frame of meson, such as , 

\ba
\label{reco2}
\vec{R}&=&\frac{(\vec{r_1}+\vec{r_2})}{2},\,\vec{r}=\vec{r_1}-\vec{r_2};\nonumber\\
\vec{p}&=&\vec{p}_{1}+\vec{p}_{2},\,\vec{q}=\frac{\vec{p}_{2}-\vec{p}_{1}}{2}
\ea

so that we can derive,
\ba
\label{reco3}
f_q(r_1,p_1)\rightarrow f_q(\lvert \vec{R}+\frac{\vec{r}}{2} \rvert,\lvert \frac{\vec{p}}{2}+\vec{q} \rvert)\,,\nonumber\\
f_{\bar{q}}(r_2,p_2)\rightarrow f_{\bar{q}}(\lvert \vec{R}-\frac{\vec{r}}{2} \rvert,\lvert \frac{\vec{p}}{2}+\vec{q} \rvert)\,,\nonumber\\
W_M(\vert \vec{r}_1-\vec{r}_2\rvert; \lvert \vec{p}_{1}-\vec{p}_{2}\rvert)\rightarrow W_M(r,q).
\ea
Here we have also assumed that $\lvert \vec{r}\rvert$ is small compared to $\lvert \vec{R}\rvert$, and thus neglected $\lvert \vec{r}\rvert$ in the quark distributions, $f_{q/\bar{q}}$. Thus we have,

\begin{widetext}
\ba
\label{reco31}
N_M=g_M\int{d^3r\,\frac{d^3R}{(2\pi)^6}}\int{\frac{d^2q\,d^2p_T}{(2\pi)^6}}\,m_{T1}.m_{T2}.f_q(\lvert \vec{R}\rvert,\lvert \frac{\vec{p}_T}{2}-\vec{q}\rvert)\,f_{\bar{q}}(\lvert \vec{R}\rvert,\lvert \frac{\vec{p}_T}{2}+\vec{q}\rvert)\,W_M(r,q)\,,
\ea
\end{widetext}
We have now,

\begin{widetext}
\ba
\label{reco4}
\centering
\frac{dN_M}{d^2p_T}=g_M\int{\frac{d^3R}{(2\pi)^3}}\int{\frac{d^2q\,d^3r}{(2\pi)^6}}\,m_{T1}.m_{T2}.f_q(\lvert \vec{R}\rvert,\lvert \frac{\vec{p}_T}{2}-\vec{q}\rvert)\,f_{\bar{q}}(\vert \vec{R}\rvert,\lvert \frac{\vec{p}_T}{2}+\vec{q}\rvert)\,W_M(r,q)\,.
\ea
\end{widetext}
where, meson transverse mass factor is given by, $\displaystyle M_T=\sqrt{p_T^2+M^2}$.

As for the Wigner function, $W_M$, we can use the following relation,
\ba
\label{reco41}
W_M(q)=\int{d^3r\,W_M(r,q)}\,.
\ea

Therefore eq.~\ref{reco4} is transformed as,
\begin{widetext}
\ba
\label{reco5}
\frac{dN_M}{d^2p_T}=g_M\int{\frac{d^3R}{(2\pi)^3}}\int{\frac{d^2q}{(2\pi)^3}}\,m_{T1}.m_{T2}.f_q(\lvert \vec{R}\rvert,\vert \frac{\vec{p}_T}{2}-\vec{q}\rvert)\,f_{\bar{q}}(\lvert \vec{R}\rvert,\lvert \frac{\vec{p}_T}{2}+\vec{q}\rvert)\,W_M(q)
\ea
\end{widetext}

To simplify our equations, we convert our momentum variable into light-cone co-ordinates, $k^\mu$ of the interacting quarks in the momentum space of the hadron as follows,

\ba
\label{reco6}
\vec{q}&=&\frac{\vec{p}}{2}- \vec{k}\,,\nonumber\\
\text{so that,} \,d^3q&=&d^3k \text{, and}\, d^3k=dk^+d^2k_\perp\,,\nonumber\\
k^\pm&=&\frac{(k_0\pm k_3)}{\sqrt{2}}\,,\nonumber\\
k_\perp^2&=&2k^{+}k^{-}-k^2\,,\nonumber\\
\text{and} \, k^+&=&x.p^+
\ea

We also assume that the partons recombining into hadrons have their momenta almost parallel to the final hadron. So $k_\perp$ can be considered to be very small compared to $k^+$ and it's dependency in the quark distribution, $f_{q/\bar{q}}$ has also been neglected. It can be shown following eq.~\ref{reco6}, the parton momentum, $k\approx x.p$, where $x$, is the momentum fraction of the final hadron's momentum, carried by its constituent quarks during recombination~\cite{Peitzmann:2005ty,GKuipers:MasterThesis}. 
Putting the above conditions into the equation, and assuming the normalization,
\ba
\label{reco7}
\int{\frac{dx\;d^2k_\perp \;p^+}{(2\pi)^3}\,W_M(x, k_\perp^2)}=1,
\ea

Finally we can write,

\begin{widetext}
\ba
\label{reco8}
\frac{dN_M}{d^2p_T}=g_M\int{\frac{d^3R}{(2\pi)^3}}\int_0^1{dx\,f_q(\lvert \vec{R}\rvert,xp_T)\,f_{\bar{q}}(\lvert \vec{R}\rvert,(1-x)p_T)\,W_M(x)}\,.
\ea
\end{widetext}

We have $d^4R=p_\mu.d\sigma^{\mu}$ along the unit normal direction, $u(R)=(1,0,0,0)$ at the freeze-out hyper-surface. 
 Similarly, for the baryons, one can derive to show,

\begin{widetext}
\ba
\label{reco9}
\frac{dN_B}{d^2p_T}=g_B\int{\frac{d^3R}{(2\pi)^3}}\int_0^1{dx_1\int_0^1{dx_2\,f_q(\lvert \vec{R}\rvert,x_1p_T)\,f_q(\lvert \vec{R}\rvert, x_2p_T)\,f_q(\lvert \vec{R}\rvert, (1-x_1-x_2)p_T)\;W_B(x_1,x_2)}}\,.
\ea
\end{widetext}

\begin{figure*}[ht]
\includegraphics[height=20em]{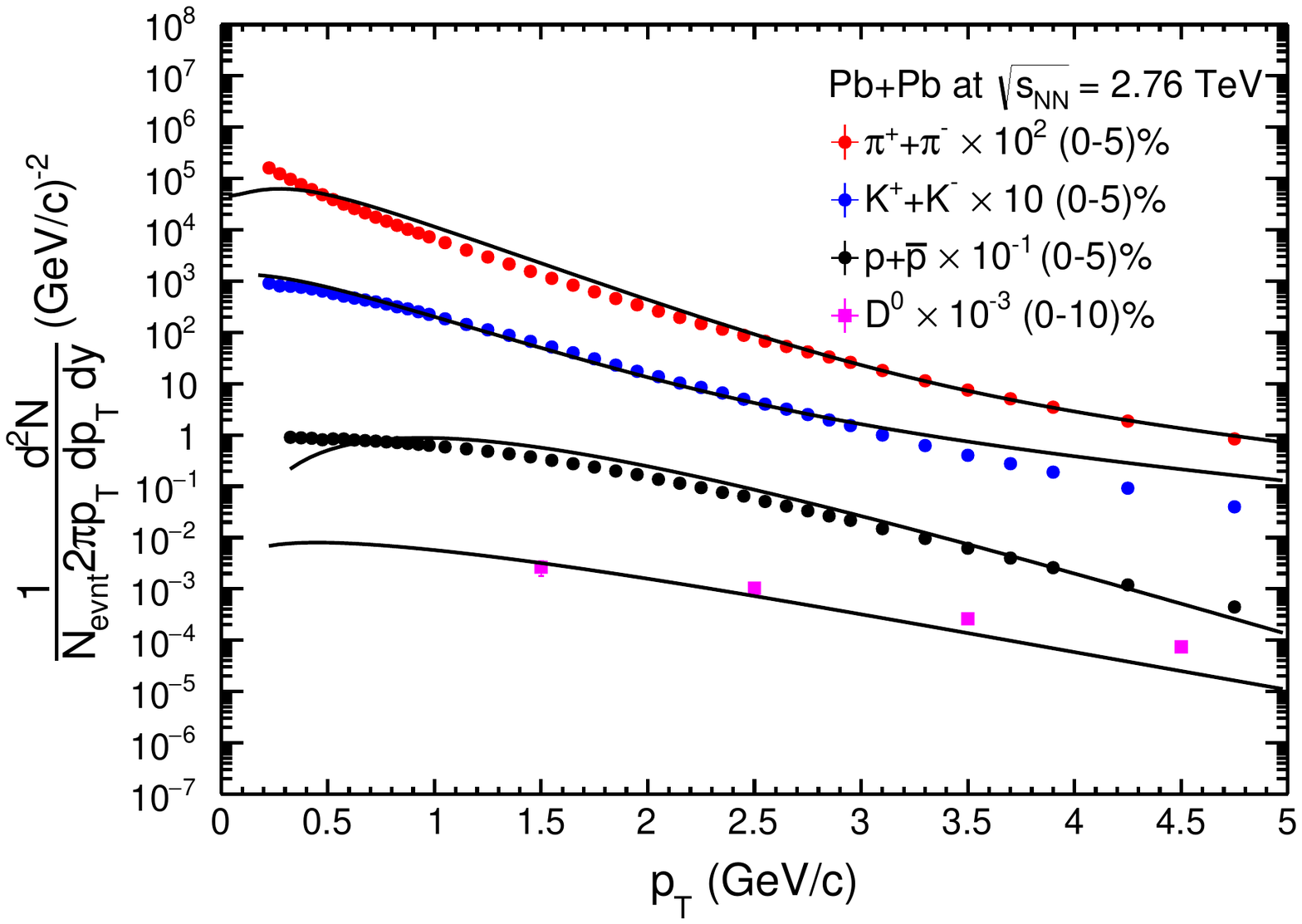}
\caption[]{(Color online) The transverse momentum spectra of ($\pi^{+}+\pi^-$), ($K^{+}+K^{-}$),($p+\bar{p}$) and $D^0$ meson versus $p_T$ for most central Pb+Pb collisions at $\sqrt{s_{NN}}$ = 2.76 TeV. Symbols are experimental data points~\cite{Abelev:2014laa,Adam:2015sza} and lines are the model results from eqs.~\ref{reco8} and ~\ref{reco9}.}
\label{ptspectra}
\end{figure*}

\begin{figure*}[ht]
\includegraphics[height=20em]{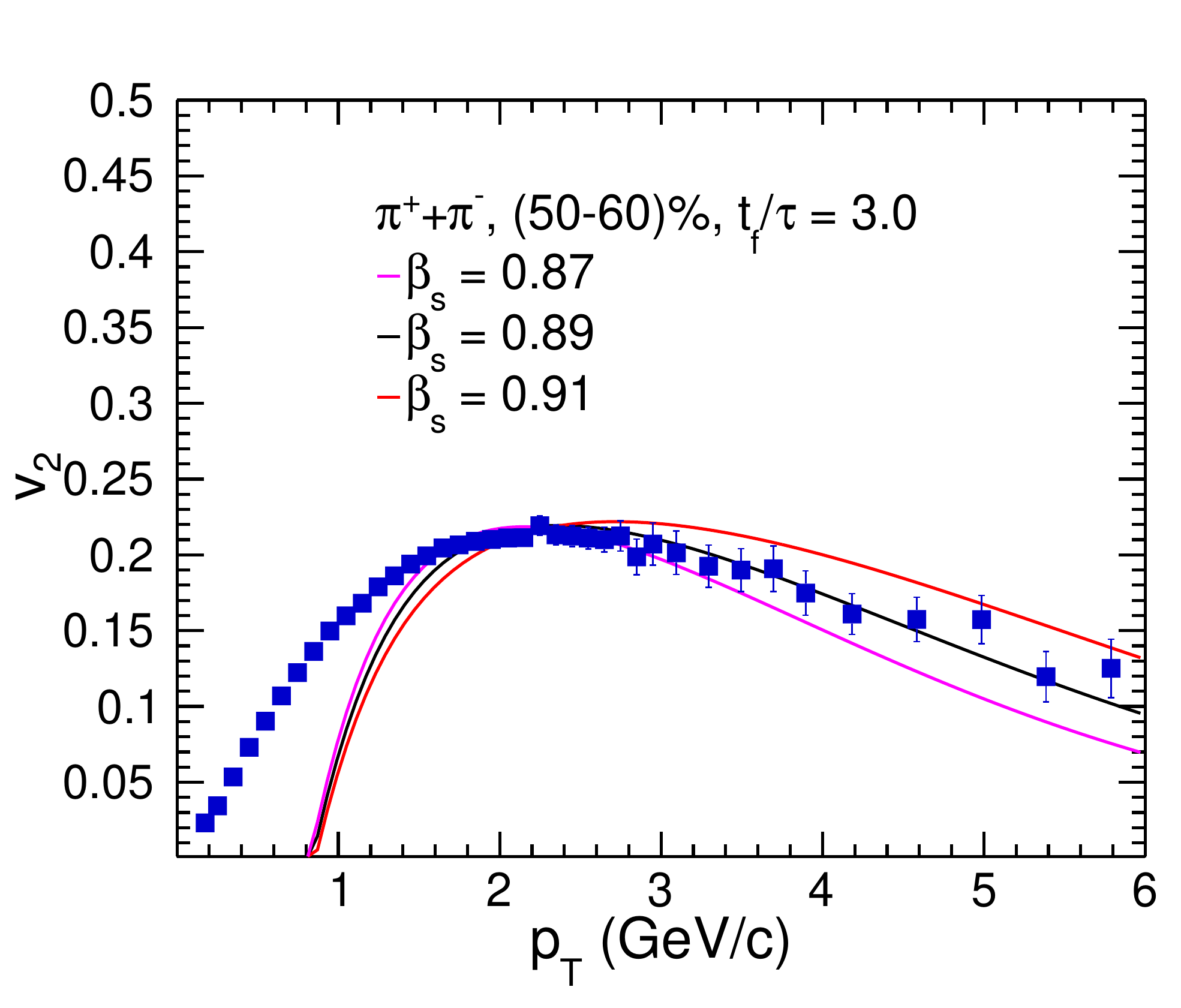}
\includegraphics[height=20em]{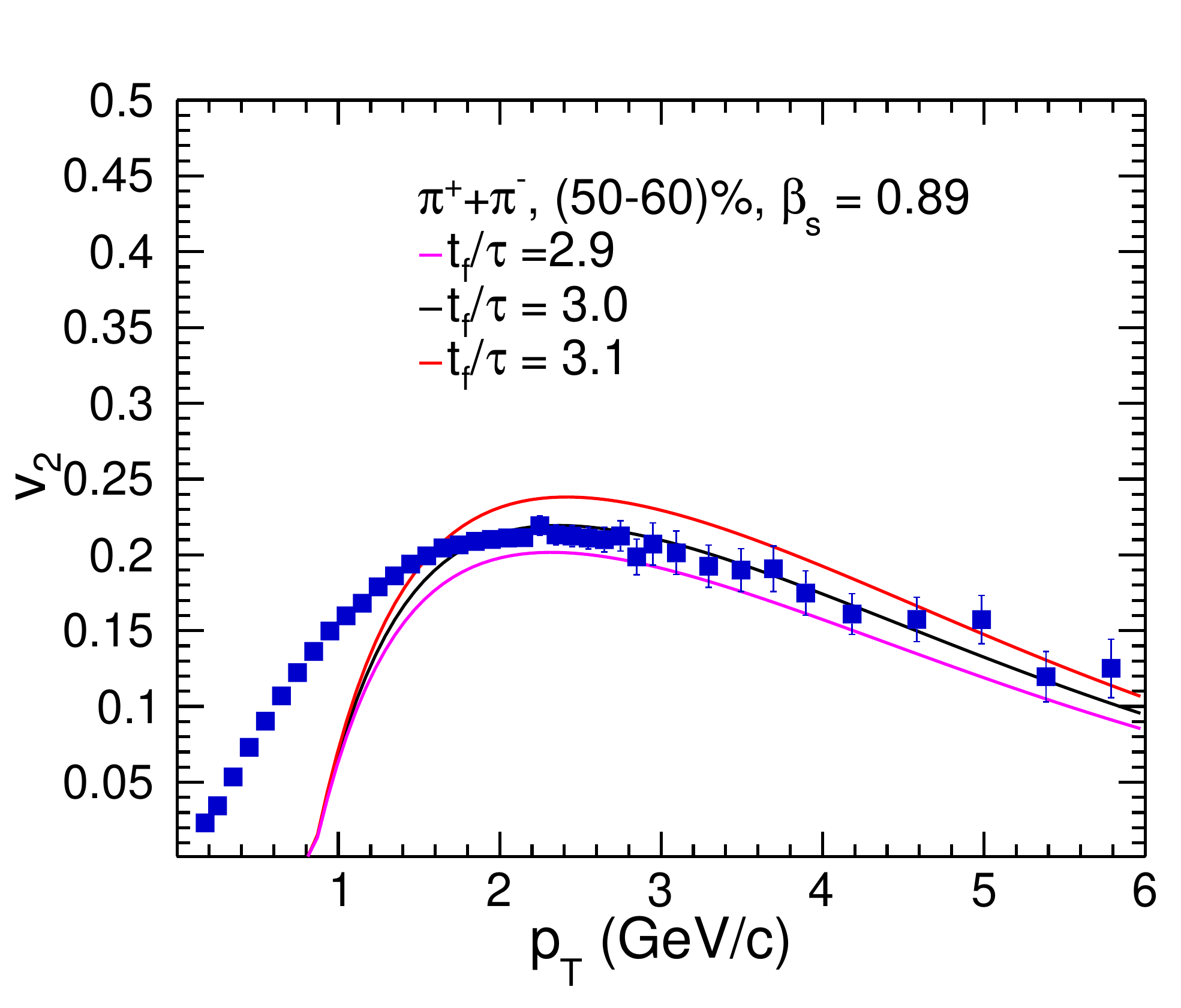}
\caption[]{(Color online) The elliptic flow ($v_2$) of ($\pi^{+}+\pi^-$) versus $p_T$ at constant $\frac{t_f}{\tau}$ and $\beta_s$ for peripheral collisions (50-60)\% at $\sqrt{s_{NN}}$ = 2.76 TeV. Symbols are experimental data points~\cite{Abelev:2014pua} and lines are the model results.}
\label{pionv2}
\end{figure*}

\begin{figure*}[ht]
\includegraphics[height=20em]{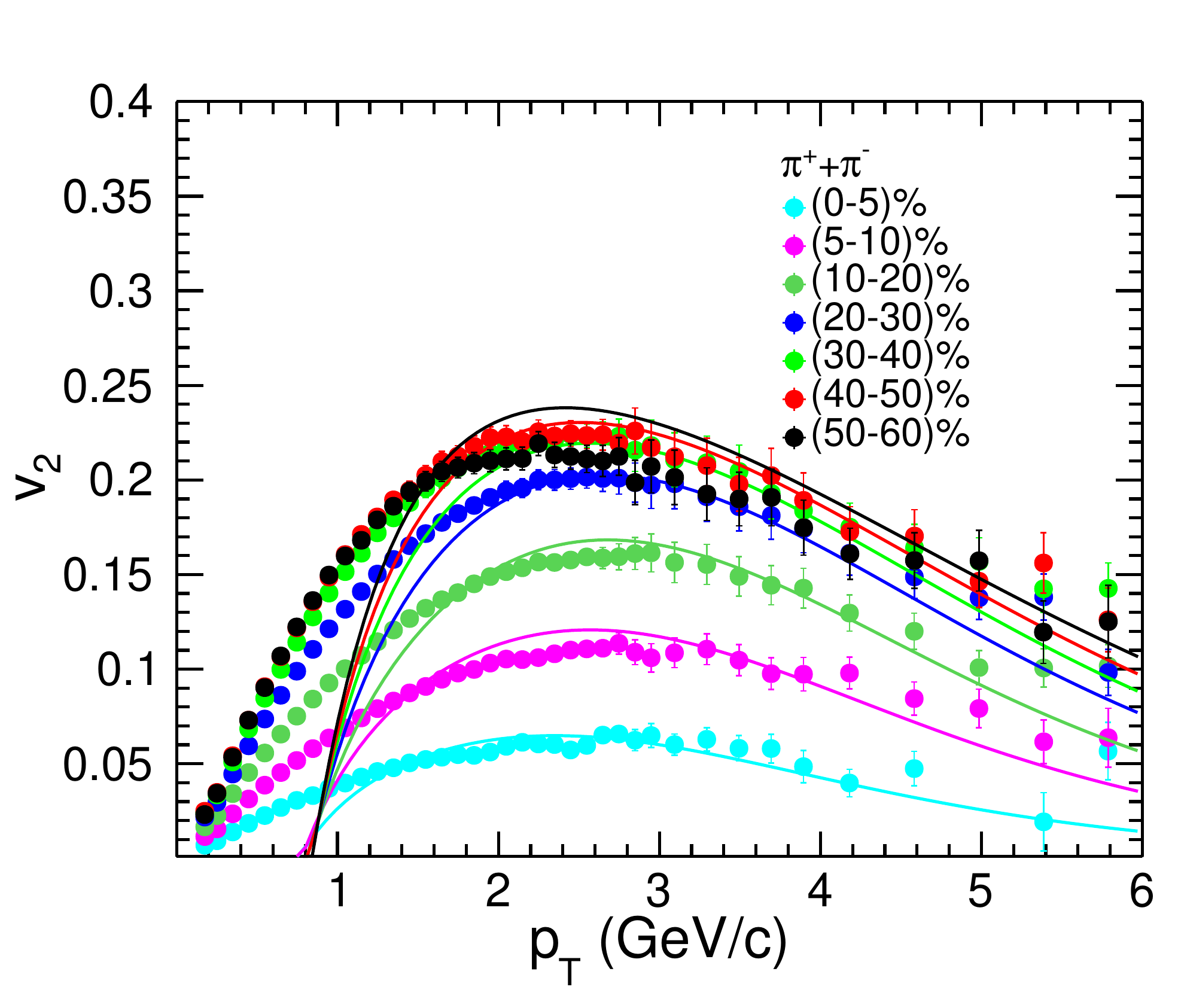}
\caption[]{(Color online) The elliptic flow ($v_2$) of ($\pi^{+}+\pi^-$) versus $p_T$ for various centralities for Pb+Pb collisions at $\sqrt{s_{NN}}$ = 2.76 TeV. Symbols are experimental data points~\cite{Abelev:2014pua} and lines are the model results.}
\label{pion}
\end{figure*}

\begin{figure*}
\includegraphics[height=17em]{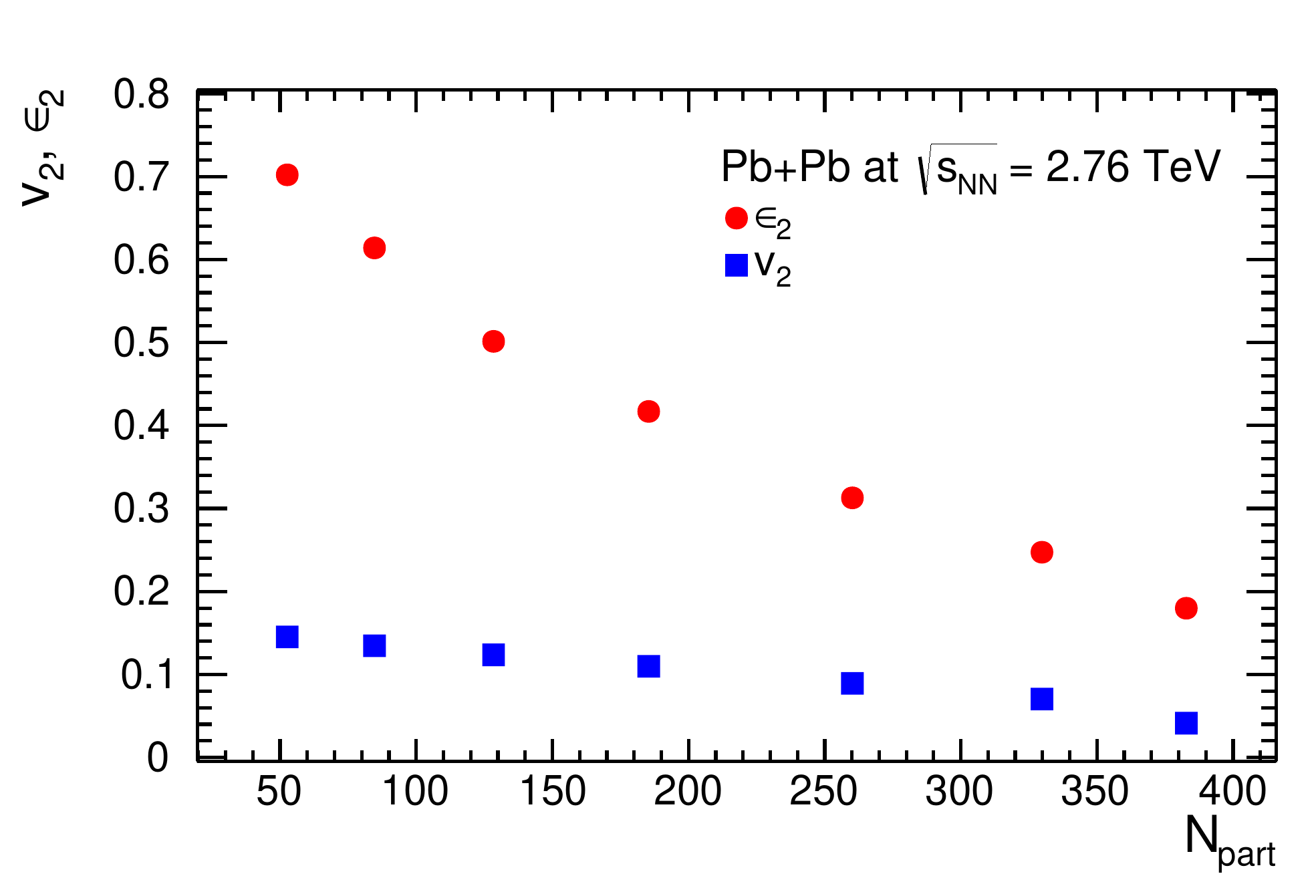}
\includegraphics[height=17em]{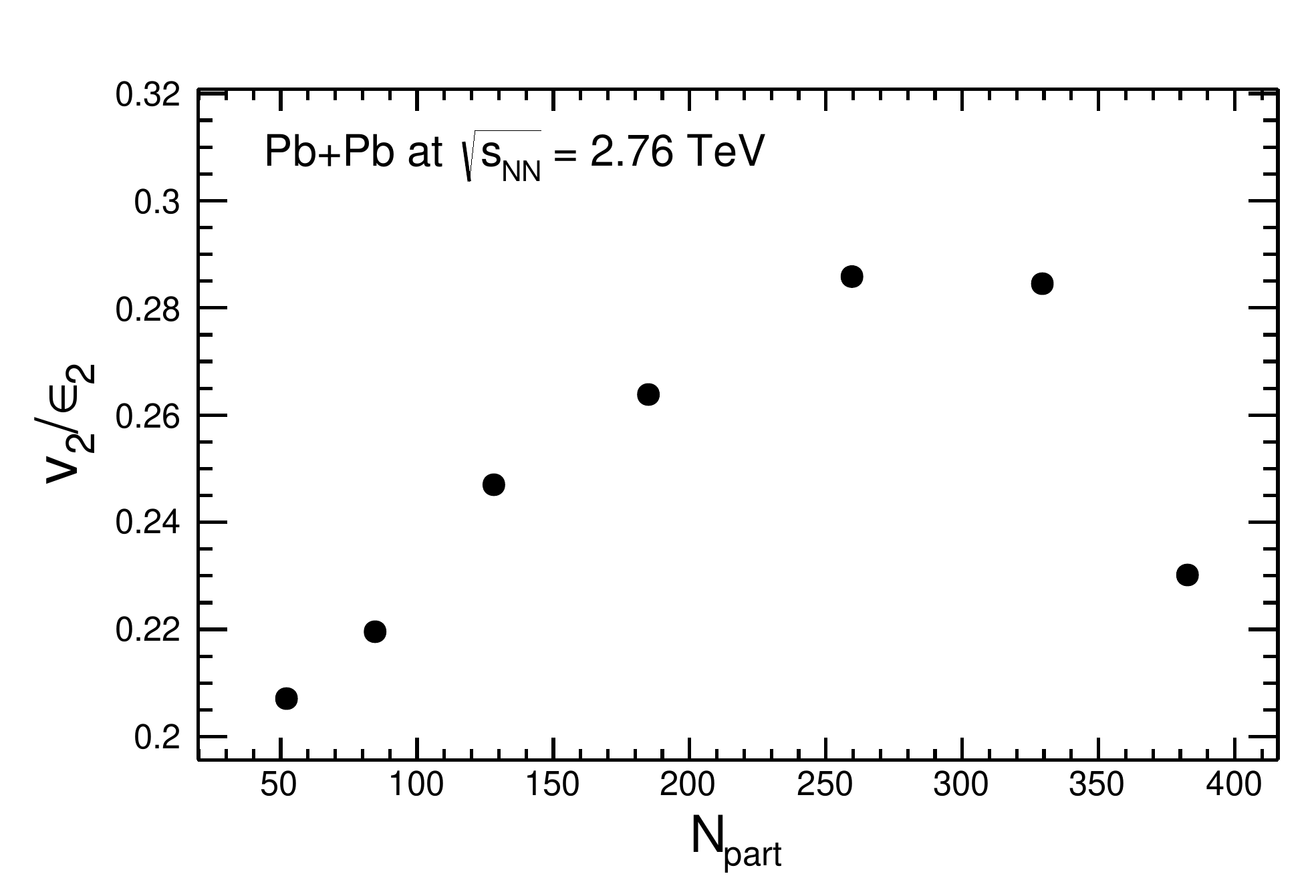}
\caption[]{(Color online) (Left)The elliptic flow ($v_2$)  and spatial anisotropy ($\epsilon_2$) of ($\pi^{+}+\pi^-$) versus $N_{part}$ for Pb+Pb collisions at $\sqrt{s_{NN}}$ = 2.76 TeV. (Right) The ratio of $v_2$ and $\epsilon_2$ of pions versus $N_{part}$.}
\label{v2e2}
\end{figure*}
To illustrate on our calculations, one may use eq.~\eqref{feq} as an example and show that
\ba
\label{reco10}
&f^{eq}_q(R,xp).f^{eq}_{\bar{q}}((1-x)p) = e^{-(p_1+p_2).u(R)/T}\,,&\nonumber\\
&(p_1+p_2).u(R)=\mu_T^M(x,p_T)\cosh\rho-p_T\sinh\rho\cos(\phi_r-\phi_p)&\nonumber\\
\ea
Thus one may calculate to show,
\begin{widetext}
\ba
\label{reco11}
\mu_T^M(x,p_T)&=&\sqrt{m_q^2+x^2p_T^2}+\sqrt{m_{\bar{q}}^2+(1-x)^2p_T^2}\,.
\ea
\end{widetext}
Similarly for the baryons, one may write,
\begin{widetext}
\ba
\label{reco12}
\mu_T^B(x_1,x_2,p_T)&=&\sqrt{m_q^2+x_1^2p_T^2}+\sqrt{m_q^2+x_2^2p_T^2}+\sqrt{m_q^2+(1-x_1-x_2)^2p_T^2}\,.\nonumber\\
\ea
\end{widetext}
We have replaced transverse mass, $m_T$, by expressions from eqs.~\eqref{reco11} and \eqref{reco12}, throughout our calculations.
We have also assumed a general Gaussian distribution as Wigner functions, $W_M$ (for mesons), $W_B$ (for baryons), which are given by,

\ba
\label{reco13}
W_M(x)&=&e^{-(x-0.5)^2/2\sigma^2_M}\,,\nonumber\\
W_B(x)&=&e^{-[(x_1-x_2)^2+(x_1+x_2-0.66)^2]/2\sigma^2_B}.
\ea

Here $2\sigma^2$ is the width of the Gaussian function, and its small values would give us the narrow Wigner function closer to being a delta function or on other hand, its larger values would give us broad convoluting function instead. The values can be chosen according to the best fit with the particle spectra. We will resume its discussion in the results section. 

Thus using eq.~\eqref{eq5}, in eqs.~\eqref{reco8} and \eqref{reco9}, we calculate $v_2$ of the final hadrons at mid-rapidity as,

\ba
v_2(p_T)\rvert_{y=0}=\frac{\int{\frac{dN_{M/B}}{d^2p_T} \times \cos(2\phi)\,d\phi}}{\int{\frac{dN_{M/B}}{d^2p_T}\,d\phi}}.
\label{v2hadrons}
\ea

\begin{table}
\begin{center}
\caption{Extracted parameters of eq.~\ref{pqcd1} at LO pQCD calculations of $p+p$ collision at $\sqrt{s_{NN}}$ = 2.76 TeV}
\begin{tabular}{ |c|c|c|c| } 
 \hline
 quark flavour & $\alpha$ & $B$(GeV) & $C(fm^4)$\\
 \hline\hline
 $u$ & 5.615 & 1.127 & 3.73376 $\times$ 10$^3$\\
 \hline 
 $\bar{u}$ & 5.999 & 1.099 & 8.73376 $\times$ 10$^2$ \\ 
 \hline
 $d$ & 5.579 & 1.434 & 3.6286 $\times$ 10$^3$\\ 
 \hline
 $\bar{d}$ & 5.953 & 1.401 & 9.1286 $\times$ 10$^2$\\ 
 \hline
 $s=\bar{s}$ & 6.523 & 1.892 & 2.6317 $\times$ 10$^2$ \\ 
 \hline
 $c=\bar{c}$ & 7.250 & 3.287 & 2.32815 \\ 
 \hline
\end{tabular}
\label{table}
\end{center}
\end{table}

\begin{figure*}
\centering
\includegraphics[height=20em]{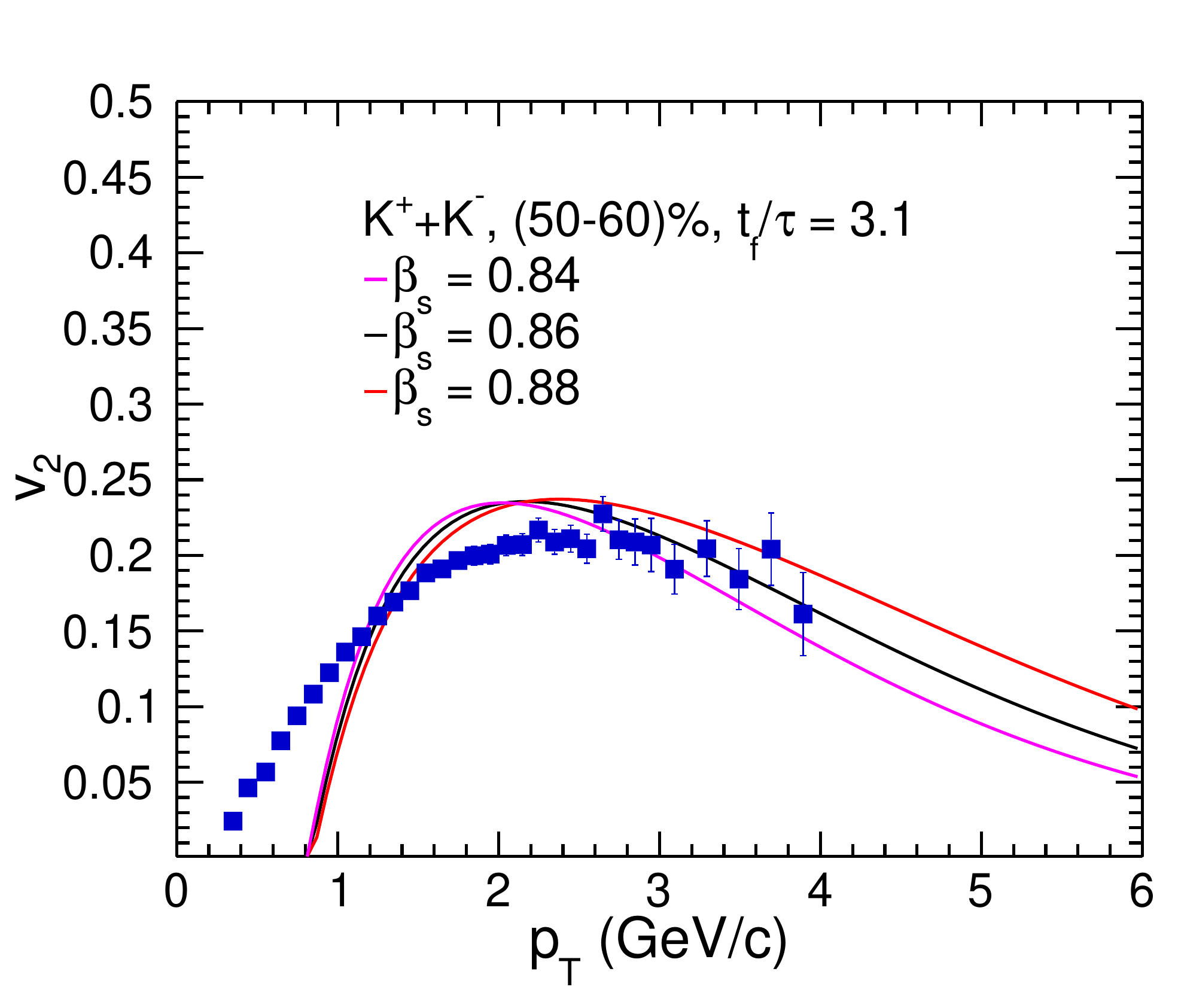}
\includegraphics[height=20em]{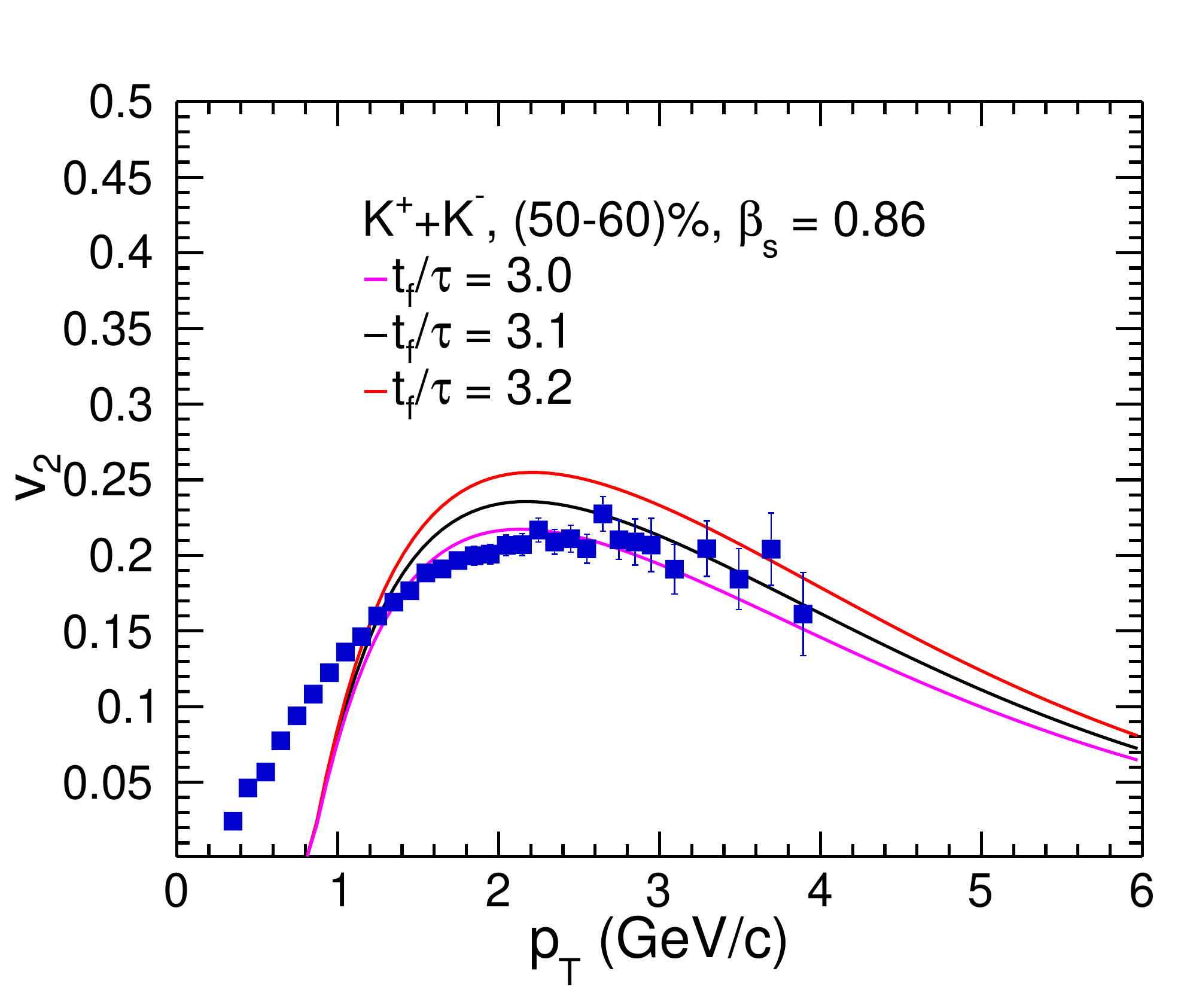}
\caption[]{(Color online) The elliptic flow ($v_2$) of ($K^{+}+K^-$) versus $p_T$ at constant $\frac{t_f}{\tau}$ and $\beta_s$ for peripheral collisions (50-60)\% at $\sqrt{s_{NN}}$ = 2.76 TeV. Symbols are experimental data points~\cite{Abelev:2014pua} and lines are the model results.}
\label{kaonv2}
\end{figure*}

\begin{figure*}
\centering
\includegraphics[height=20em]{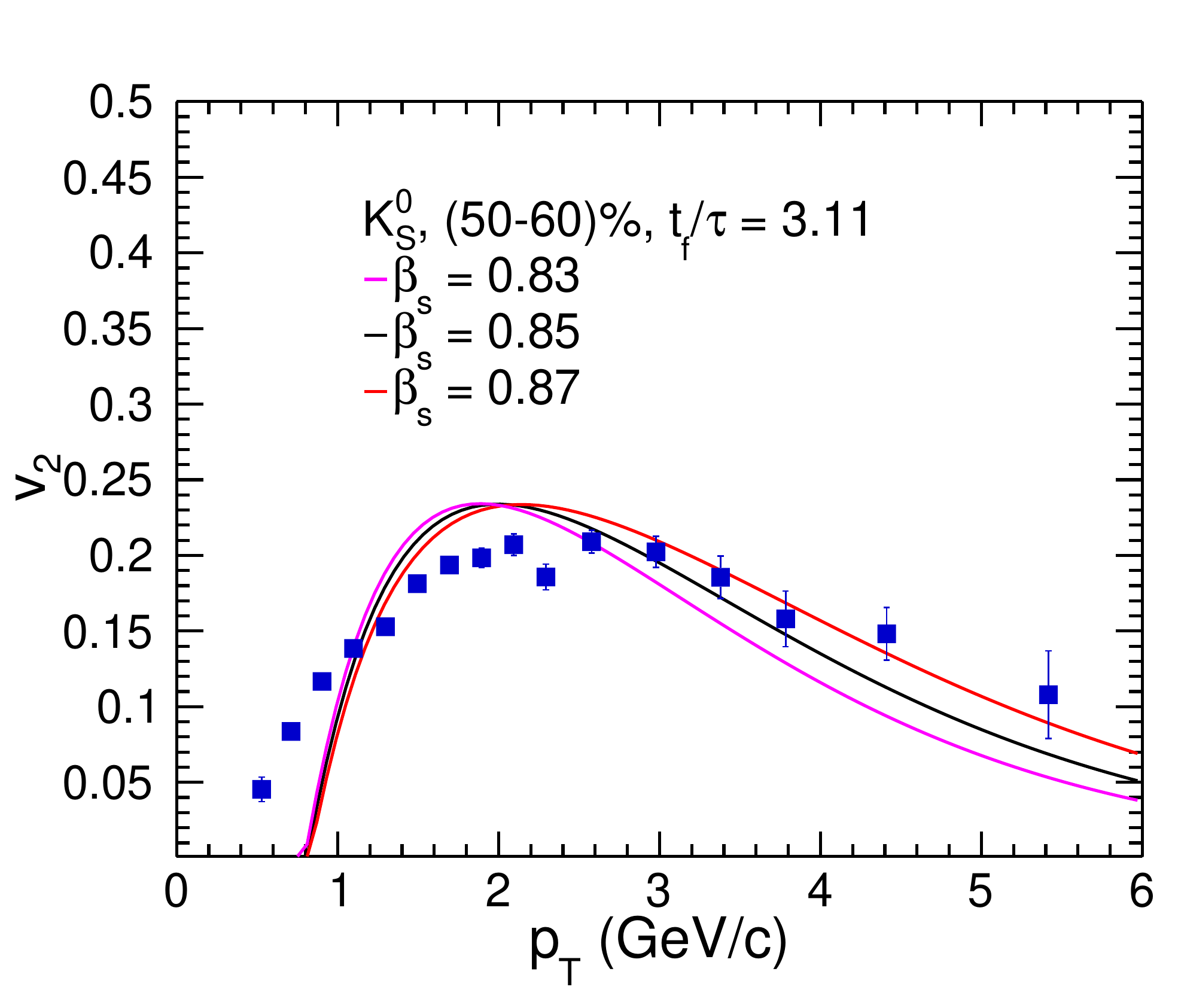}
\includegraphics[height=20em]{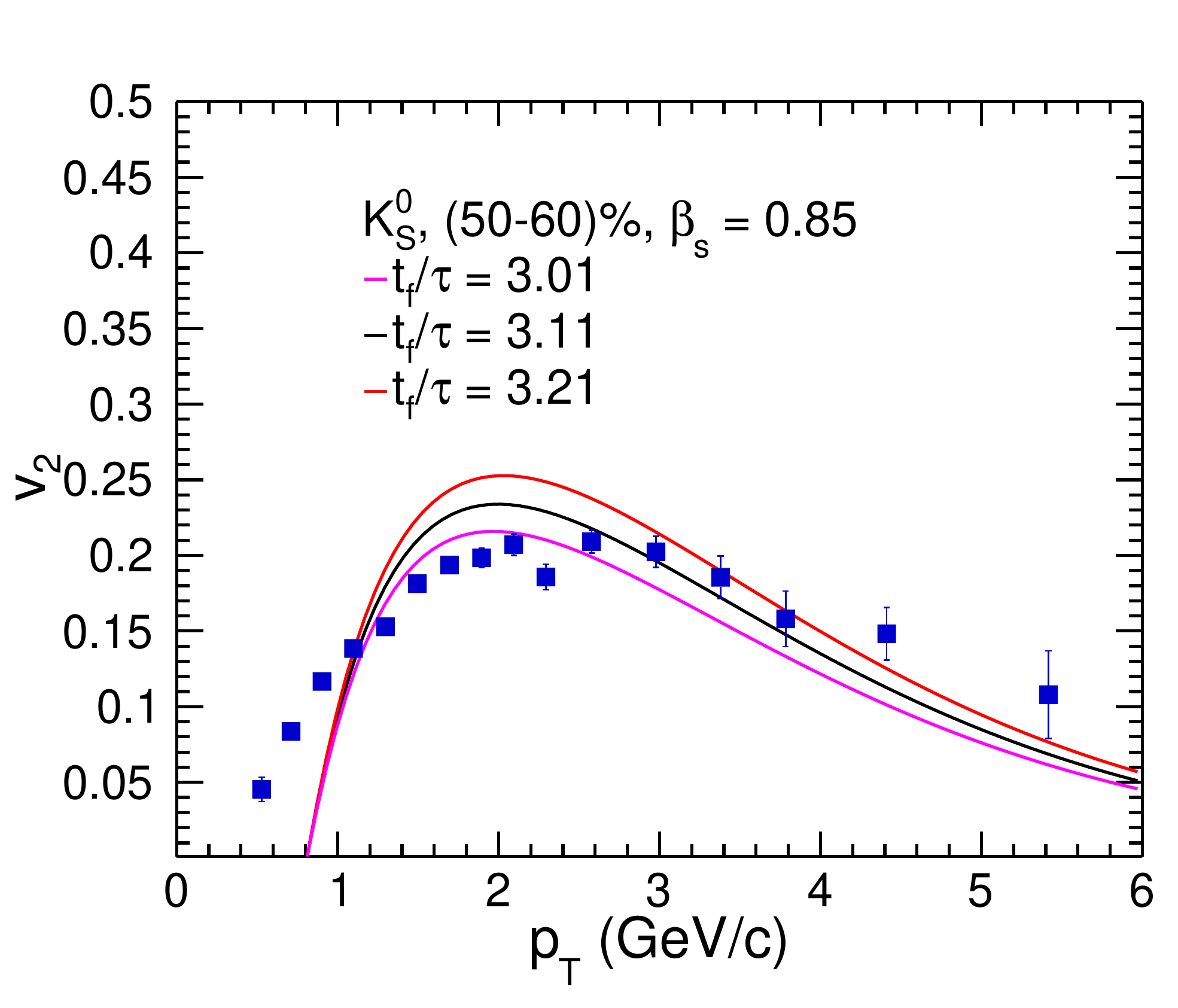}
\caption[]{(Color online) The elliptic flow ($v_2$) of $K_S^0$ versus $p_T$ at constant $\frac{t_f}{\tau}$ and $\beta_s$ for peripheral collisions (50-60)\% at $\sqrt{s_{NN}}$ = 2.76 TeV. Symbols are experimental data points~\cite{Abelev:2014pua} and lines are the model results.}
\label{kshortv2}
\end{figure*}

\begin{figure*}
\centering
\includegraphics[height=20em]{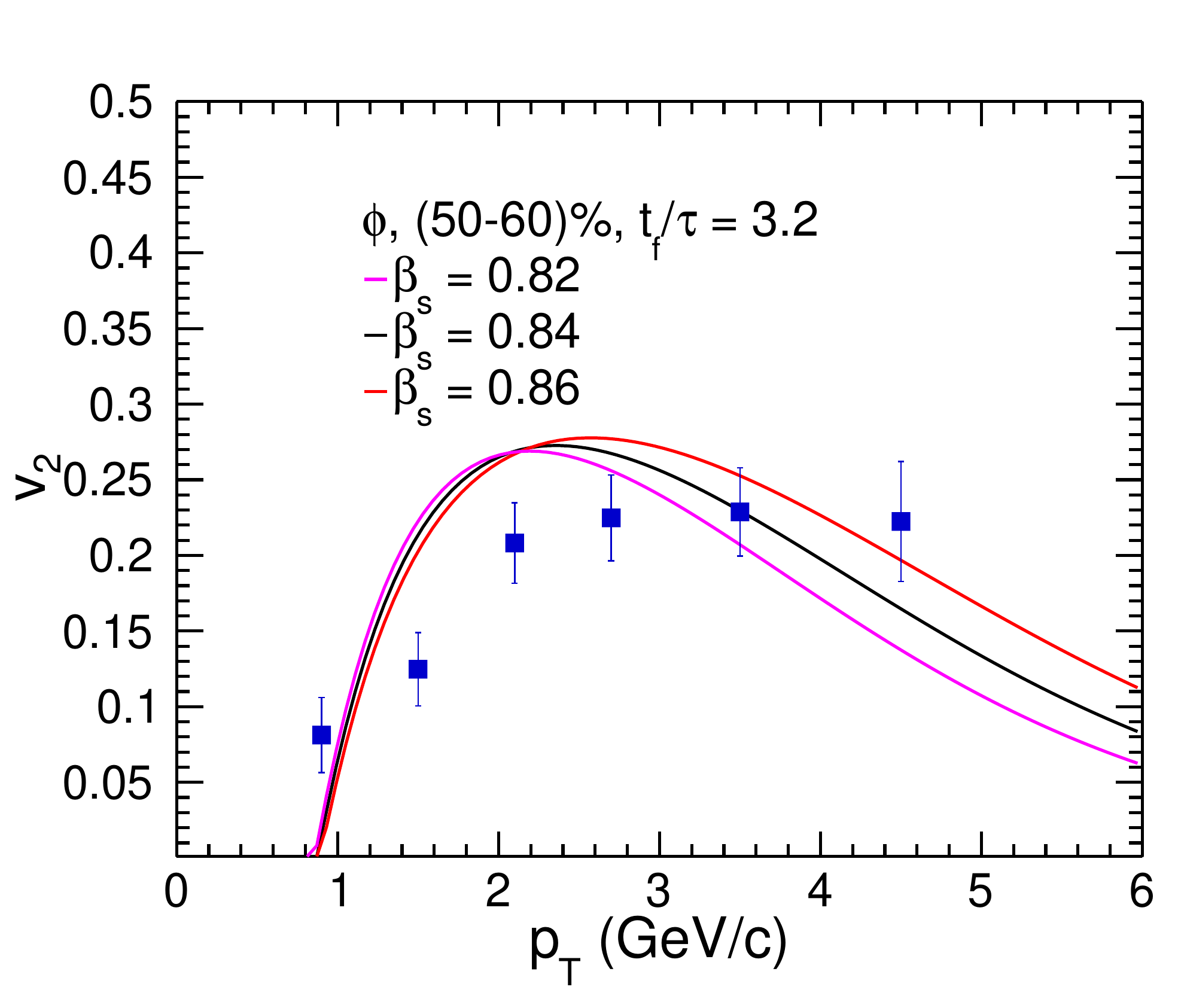}
\includegraphics[height=20em]{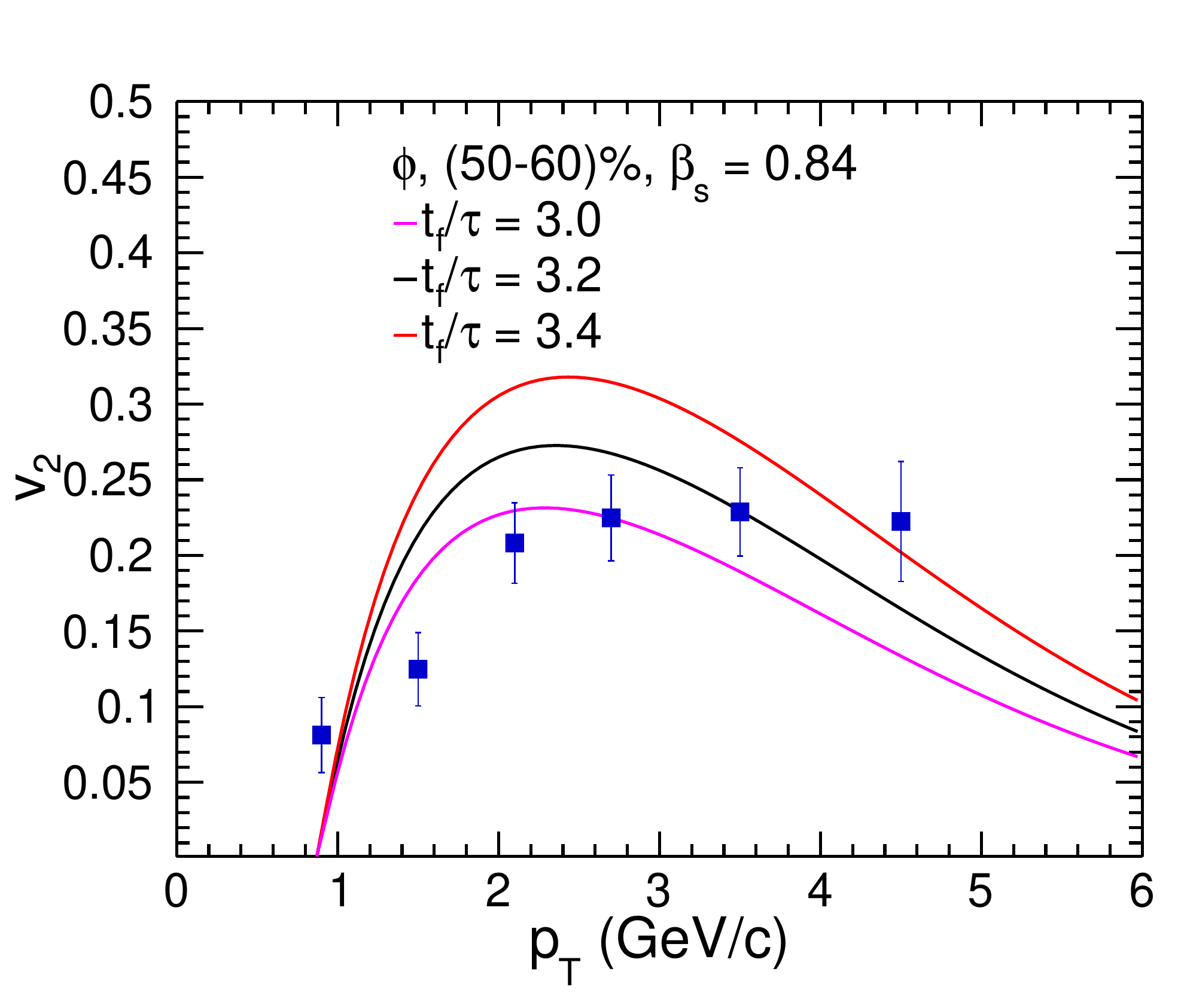}
\caption[]{(Color online) The elliptic flow ($v_2$) of phi-meson ($\phi$) versus $p_T$ at constant $\frac{t_f}{\tau}$ and $\beta_s$ for peripheral collisions (50-60)\% at $\sqrt{s_{NN}}$ = 2.76 TeV. Symbols are experimental data points~\cite{Abelev:2014pua} and lines are the model results.}
\label{phiv2}
\end{figure*}

\section{Results and Discussions}
\label{results}

\begin{figure*}
\centering
\includegraphics[height=20em]{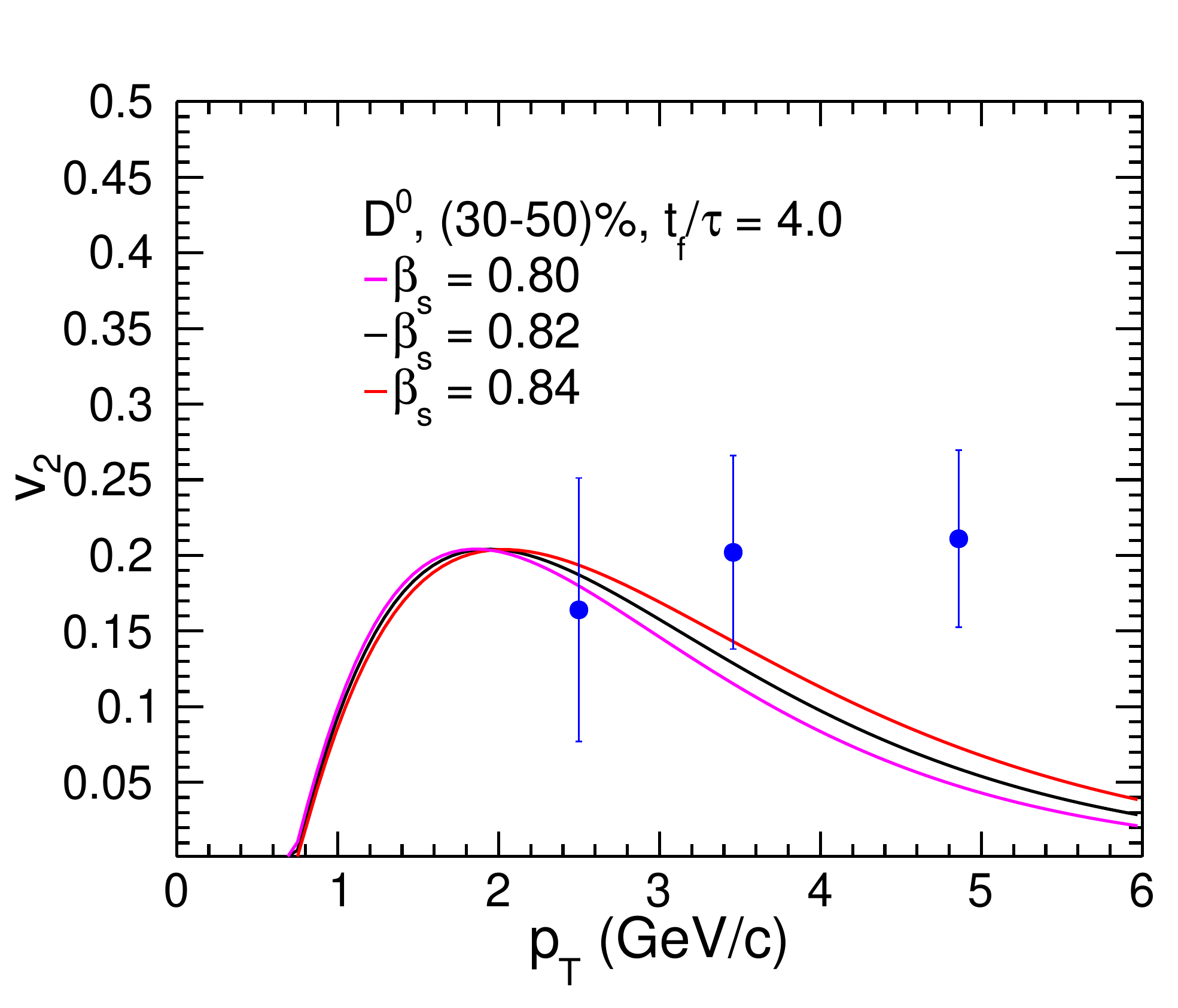}
\includegraphics[height=20em]{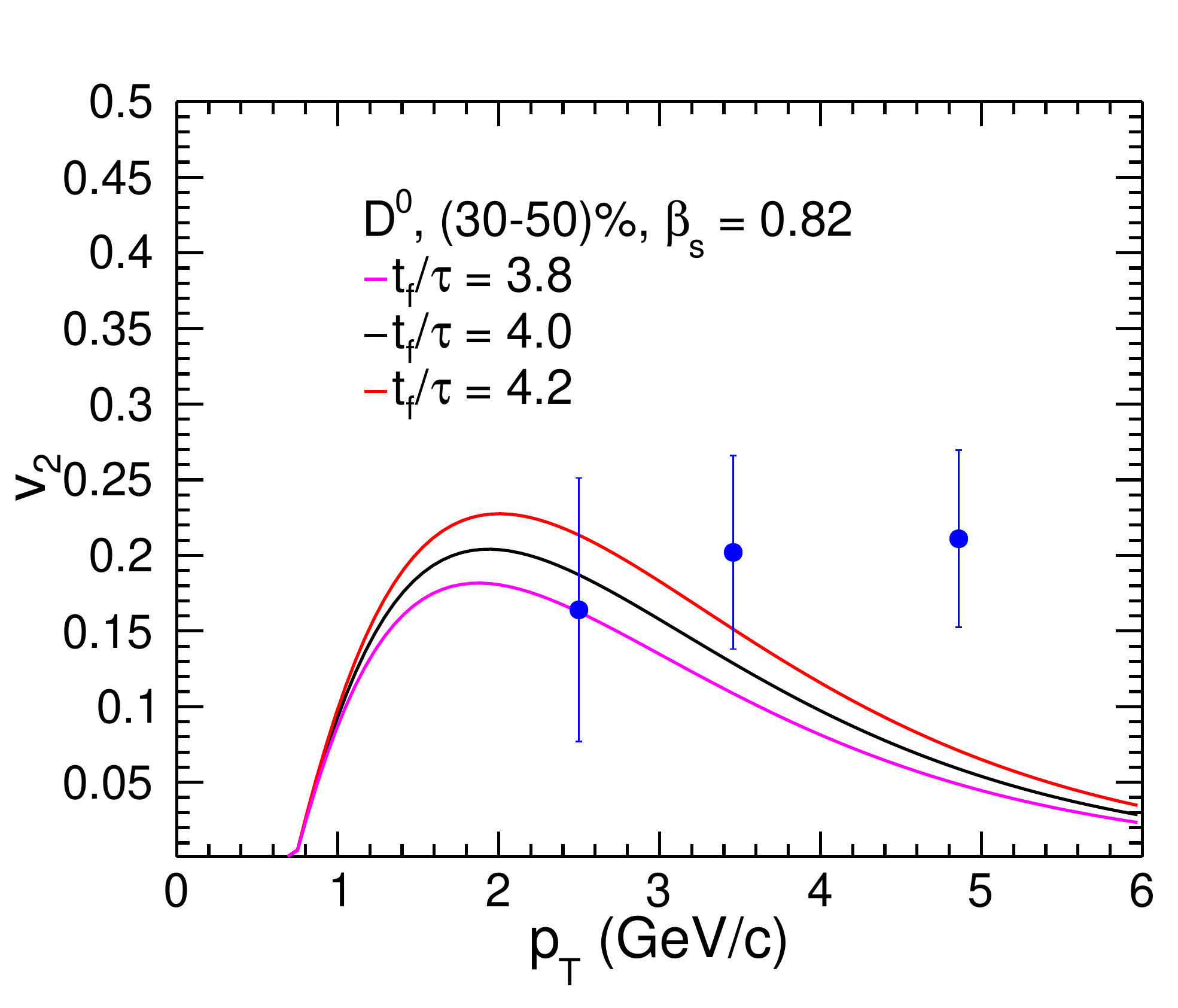}
\caption[]{(Color online) The elliptic flow ($v_2$) of D meson versus $p_T$ at constant $\frac{t_f}{\tau}$ and $\beta_s$ for peripheral collisions (30-50)\% at $\sqrt{s_{NN}}$ = 2.76 TeV. Symbols are experimental data points~\cite{Abelev:2013lca} and lines are the model results.}
\label{Dmesonv2}
\end{figure*}

We would like to re-iterate that in the current work, using BTE in RTA we have transported quarks of various flavors ($u,d,s,$ and $c$) produced promptly (assumed to be out of equilibrium) from initial gluon fusion, to thermalization and freeze-out time and recombined them into hadrons using coalescence mechanism. We have neglected the decay contributions to final hadron spectra as well as effects due to hadronic interactions. We have tried to extract the correlation between parameters such as radial flow,$\displaystyle\beta_s$, and ratio of freeze-out time and relaxation/thermalization time, $\displaystyle t_f/\tau$, We have presented the results on the elliptic flow ($v_2$) of various identified hadrons like pions, kaons, protons, D meson, lambda etc. for different centralities of Pb+Pb collision at $\sqrt{s_{NN}}$ = 2.76 TeV. While analysing the data, we have kept the freeze-out temperature ($T_f$) for the hadrons at 0.095 GeV for the most central collisions (0-5)\% and 0.11GeV for most peripheral collisions (50-60)\%~\cite{Abelev:2013vea}. We assume that the value of $T_f$ is smaller for the central collisions in comparison to the peripheral collisions. The above assumption on freeze-out temperature is based on the fact that the freeze-out in peripheral collisions occurs quicker than in the most central collisions~\cite{Abelev:2013vea}. The dependence of identified hadrons' $\displaystyle v_2$ are studied by varying two parameters, $\displaystyle\beta_s$ and $\displaystyle t_f/\tau$, using eq.~\ref{v2hadrons}.  Based on the closest explanation of the data, we have kept the width of the Wigner function, $2\sigma^2$ fixed at 0.0009 for mesons and 0.04 for baryons in our calculations. As discussed earlier that the constituent quarks for recombination process occupy a close phase space, we needed a narrow Gaussian function and not a delta function so as to avoid the collinear divergences as well as satisfy the above condition. We haven't included flow from hadronic medium as they are most visible for particles below p$_T$ $<$ 2 GeV~\cite{Nonaka:2003ew,He:2010vw} where our results focus on p$_T$ $\geq$ 2 GeV.

In fig.~\ref{ptspectra}, we have shown $p_T$ spectra of various charged hadrons in the most central collisions of Pb+Pb at $\sqrt{s_{NN}}$ = 2.76 TeV. The coalescence method is employed to form hadrons from quarks at the freeze-out surface. The resulting transverse momentum distributions are then drawn and compared with the experimental data from ALICE@CERN~\cite{Abelev:2014pua,Adam:2015sza,Abelev:2013lca}. It is found that, the discussed model in the above section explains the experimental data in the moderate $p_T $ region.

In fig.~\ref{pionv2}, we have shown $v_2$ of ($\pi^++\pi^-$). The left plot shows the variation of $v_2$ with $p_T$ for different surface velocity parameter, $\displaystyle \beta_s$, while the right plot shows for different $\displaystyle t_f/\tau$. Three different values of $\beta_s$, keeping $t_f/\tau$ fixed are taken and vice versa.  Generally speaking, our theoretical results match with the experimental data within errors, from the mid-$p_T$ region to the max. $p_T$ shown. However, the model fails to explain the data for $p_T<$ 1.0 GeV/c. The reason may be due to the absence of pions from decays of resonances~\cite{Greco:2004ex}. Pions also stand out as an example that shows coalescence picture should work mostly in the mid-$p_T$ region. 

In fig.~\ref{pion}, the elliptic flow of pions ($\pi^++\pi^-$) is presented as a function of $p_T$ for various centralities at $\sqrt{s_{NN}}$ = 2.76 TeV for Pb+Pb collisions. Symbols are the experimental data and lines are model results. Here freeze-out temperatures ($T_f$) are taken smaller for most central collision than to peripheral collisions. The model results are found to explain the data qualitatively beyond $p_T$ = 1 GeV/c for all the centralities within error-bar. However, the quark-coalescence mechanism is not able to explain the data below $p_T$ = 1 GeV/c. Experimentally, it is found that $v_2$ for (50-60)\% appears to be inverse in order compared to (40-50)\% due to statistical fluctuations. However, the model follows the expected trend of higher $v_2$ for higher centralities.

In the left panel of fig.~\ref{v2e2}, we have shown elliptic flow or azimuthal anisotropy $v_2$ and spatial anisotropy $\epsilon_2$ of the pions versus $N_{part}$. $\epsilon_2$ is generally defined in terms of spatial coordinates $(x,y)$ of participants nucleons in the transverse plane. It can be written as:
\ba
\epsilon_2=\frac{\langle x^2-y^2\rangle}{\langle x^2+y^2\rangle}.
\ea
In this paper, Glauber-MC formalism~\cite{Miller:2007ri} has been employed to calculate $\epsilon_2$. Both $v_2$ and $\epsilon_2$ decrease with $N_{part}$, which is expected. In the right panel of fig.~\ref{v2e2}, we show the ratio of $v_2$ and $\epsilon_2$ vs. $N_{part}$ or centrality. We find that the ratio tends to increase  towards central collisions but drops suddenly for most central. This ratio approximately shows the strength of anisotropy developed as we move towards central collisions and may indicate the extent of collectivity undertaken by the bulk of the partons within quark gluon plasma. However, the sudden drop in this ratio at the most central will be investigated further in our future reports.

\begin{figure*}
\centering
\includegraphics[height=20em]{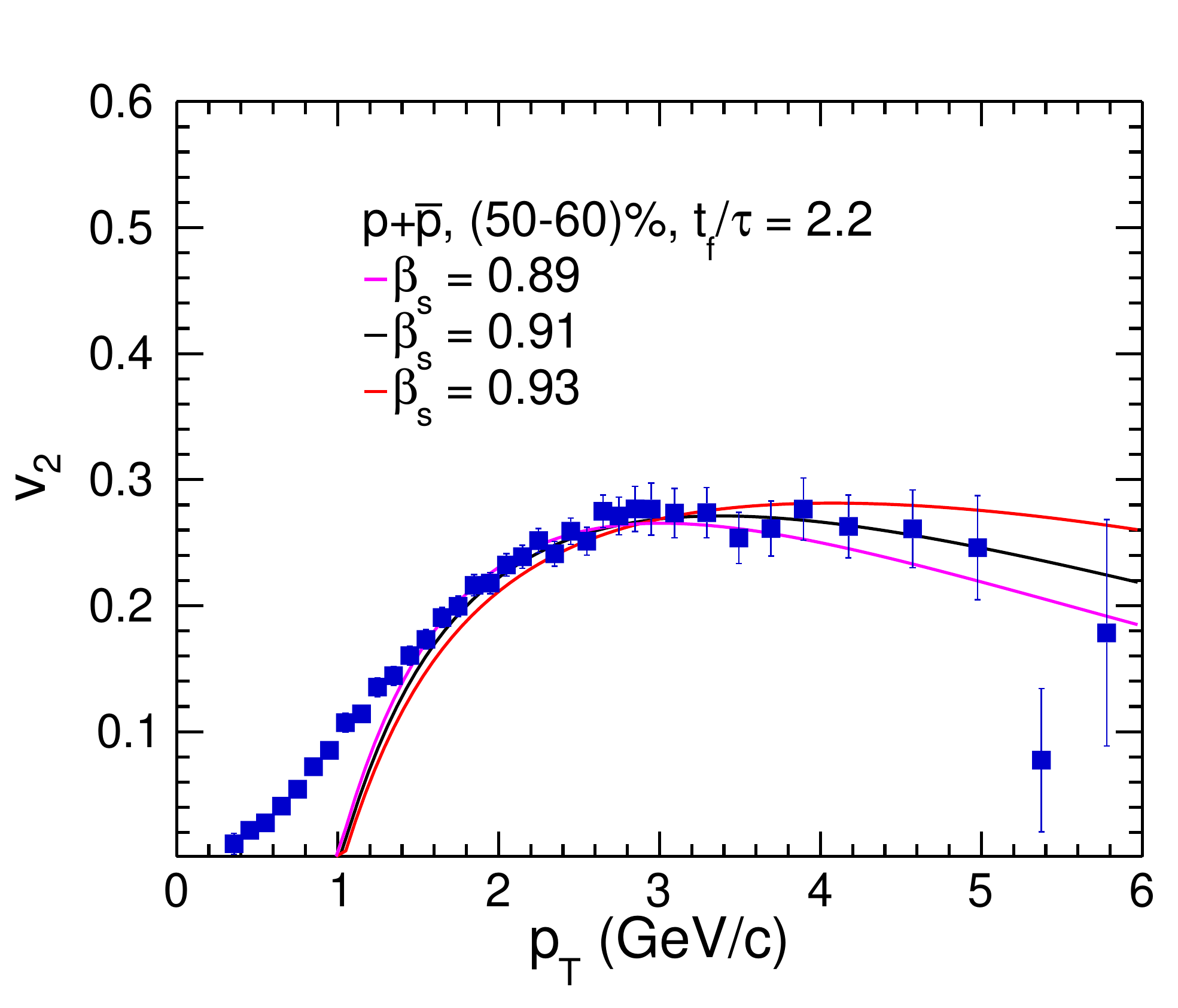}
\includegraphics[height=20em]{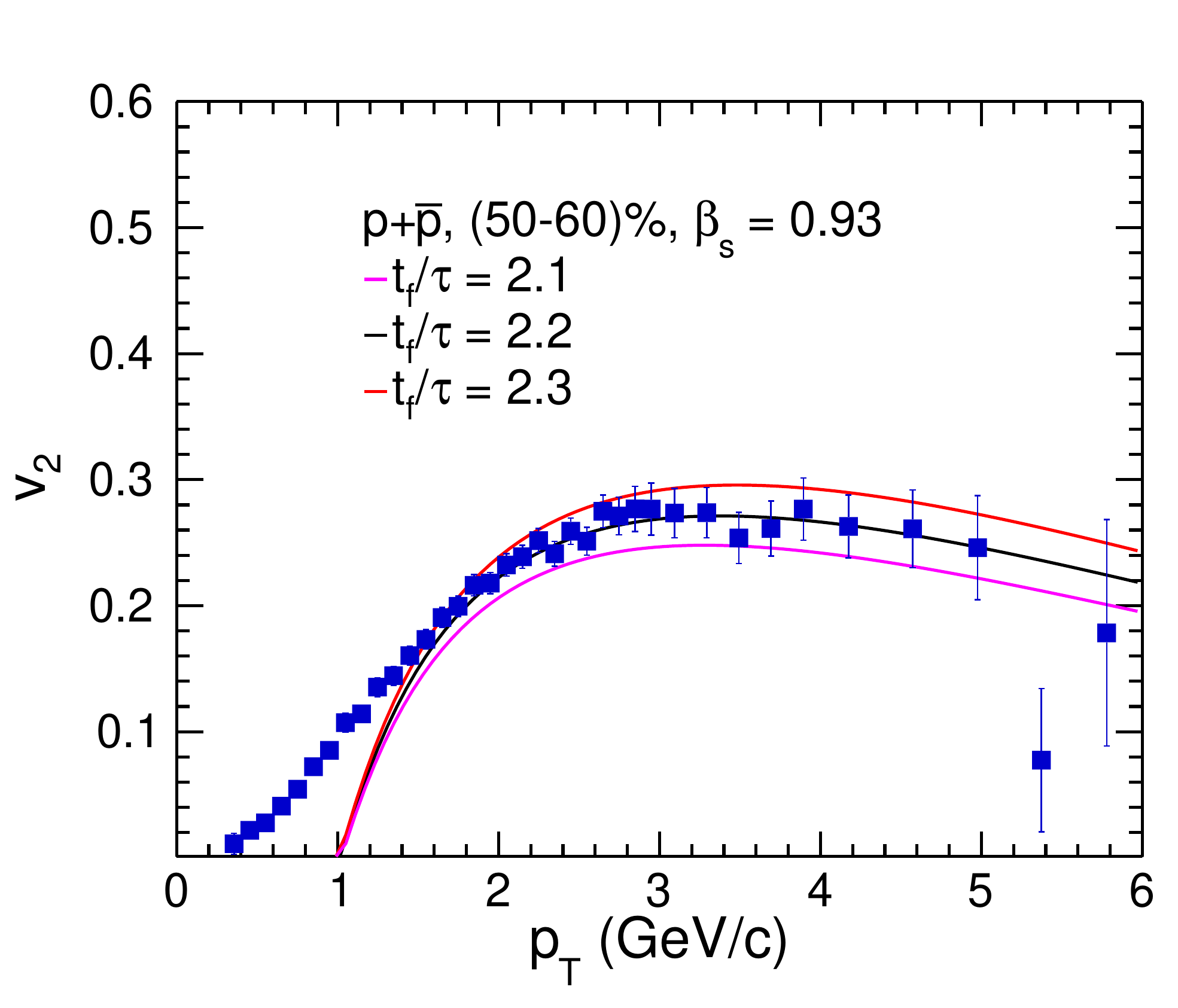}
\caption[]{(Color online) The elliptic flow ($v_2$) of ($p+\bar{p}$) versus $p_T$ at constant $\frac{t_f}{\tau}$ and $\beta_s$ for peripheral collisions (50-60)\% at $\sqrt{s_{NN}}$ = 2.76 TeV. Symbols are experimental data points~\cite{Abelev:2014pua} and lines are the model results.}
\label{protonv2}
\end{figure*}

In fig.~\ref{kaonv2}, we have presented the variations of $v_2$ of Kaons, ($K^{+}+K^-$) with $p_T$ for 50-60 \% centrality. The left panel is $v_2$ versus $p_T$ for different $\displaystyle\beta_s$ at constant $t_f/\tau$, while the right panel shows $v_2$ versus $p_T$ for different $\displaystyle t_f/\tau$ at constant $\displaystyle\beta_s$. Three different values of $\beta_s$, keeping $t_f/\tau$ fixed are taken and vice versa. The theoretical curves tend to overestimates the data although it gives a consistent explanation as to the nature of shape of Kaons $v_2$ shown by the experimental data. Also, the plot on the left side shows the theoretical lines cross each other for the different values of $\beta_s$, which shows greater sensitivity of $v_2$ on the surface velocity of the fireball. The theoretical line is quite close to experimental points at low $p_T$ which shows that large  mass should have less contribution from resonance decays.

In fig.~\ref{kshortv2}, we have shown $v_2$ of $K$-short ($K_S$). The left plot shows the variation of $v_2$ with $p_T$ taking various values of $\displaystyle\beta_s$. The right plot of the figure represents the variation of $v_2$ with $p_T$ taking different values of $\displaystyle t_f/\tau$. Three different values of $\beta_s$, keeping $t_f/\tau$ fixed are taken and vice versa. $K_S$ is a little heavier than Kaons, which is why the $\displaystyle\frac{t_f}{\tau}$ and $\beta_{s}$ values are almost similar in both the cases. Similarly, the theoretical curve tends to overestimate the data up to $p_T$= 3 GeV/c. However, theoretical curve shows a gradually increasing trend and slopes down smoothly at high $p_T$.

Fig.~\ref{phiv2} represents the variations of $v_2$ with respect to $p_T$ of phi,$\phi$. The left plot shows the variation of $v_2$ for different $\displaystyle\beta_s$ at constant $\displaystyle t_f/\tau$, while the right plot shows the variation of $v_2$ with parameter $\displaystyle t_f/\tau$ keeping $\displaystyle\beta_s$ constant. Three different values of $\beta_s$, keeping $t_f/\tau$ fixed are taken and vice versa. Phi meson's results show a gradual rise in the values of $v_2$ with increase in $p_T$ as shown in the plot. Although, the data points show a very small variation after $p_T >$ 3 GeV/c, the theoretical curves drop smoothly and continues to do so at $p_T$= 6.0 GeV/c.

In fig.~\ref{Dmesonv2}, the elliptic flow of D meson is presented as a function of $p_T$ for centrality 30-50\% at $\sqrt{s_{NN}}$ = 2.76 TeV Pb+Pb collisions. The left panel is shown for various $\beta_s$ at constant $t_f/\tau$. The model shows rise in $v_2$ for $p_T<$ 3 GeV/c and falls smoothly afterwards. The data points show almost a constant $v_2$ value and also number of data points are small to be explained satisfactorily by our model. The right plot is $v_2$ of D meson for different values of $t_f/\tau$ keeping $\beta_s$ constant.

In fig.~\ref{protonv2}, we have shown the variations of elliptic flow of $p+\bar{p}$  with respect to $p_T$ for 50-60\% centrality of Pb+Pb collisions at $\sqrt{s_{NN}}$ = 2.76 TeV. In the left hand side of the figure, we show the $v_2$ for different values of $\beta_s$ keeping $t_f/\tau$ fixed. It is found that the model results explain the experimental data qualitatively above $p_T$ = 1 GeV/c for $\beta_s$ = 0.9. The right hand side of the figure is the results for various $t_f/\tau$ at constant $\beta_s$. Again there is a good agreement between the model calculations and experimental data above $p_T$ = 1 GeV/c for $t_f/\tau$ = 2.2.

\begin{figure*}
\centering
\includegraphics[height=20em]{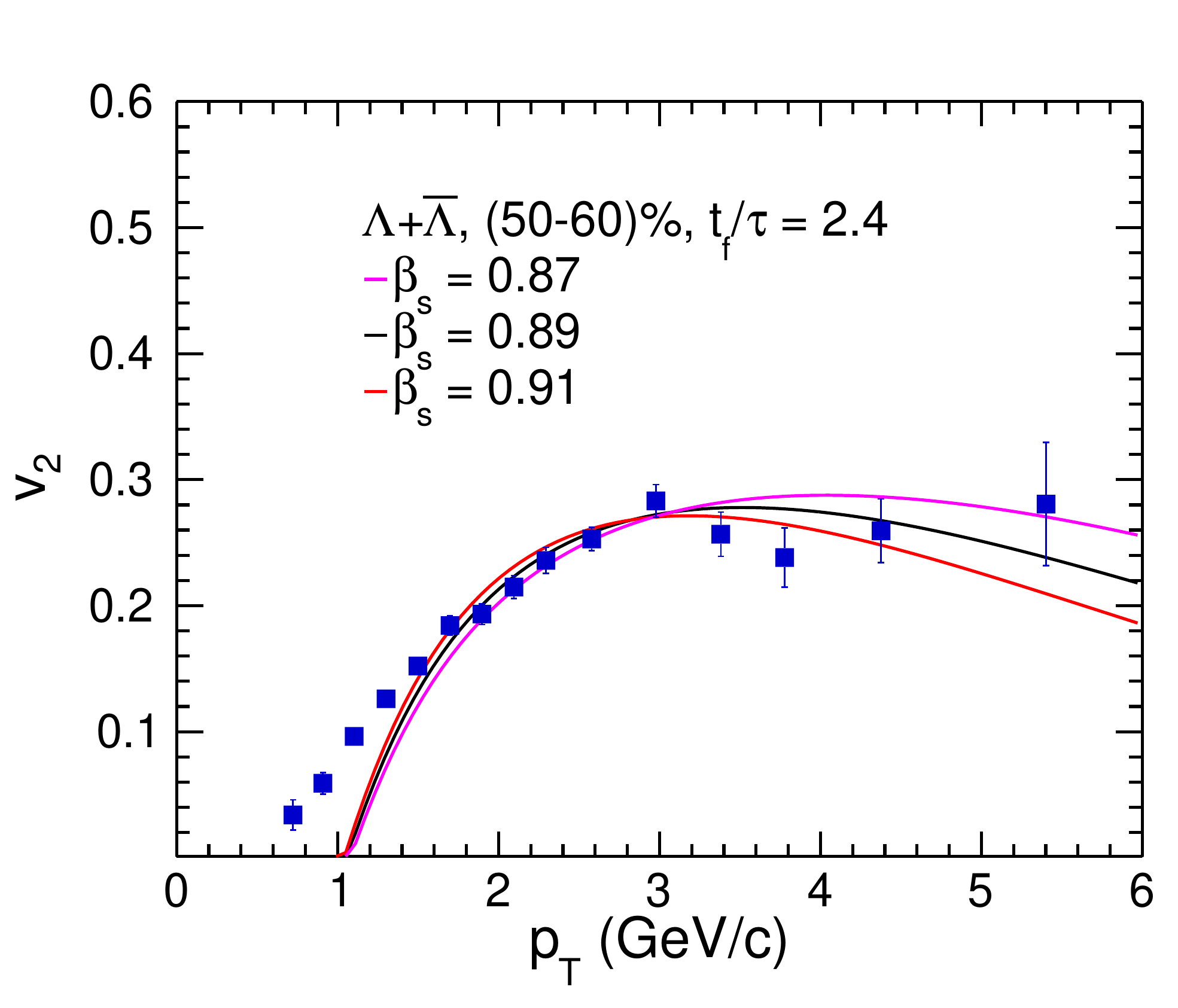}
\includegraphics[height=20em]{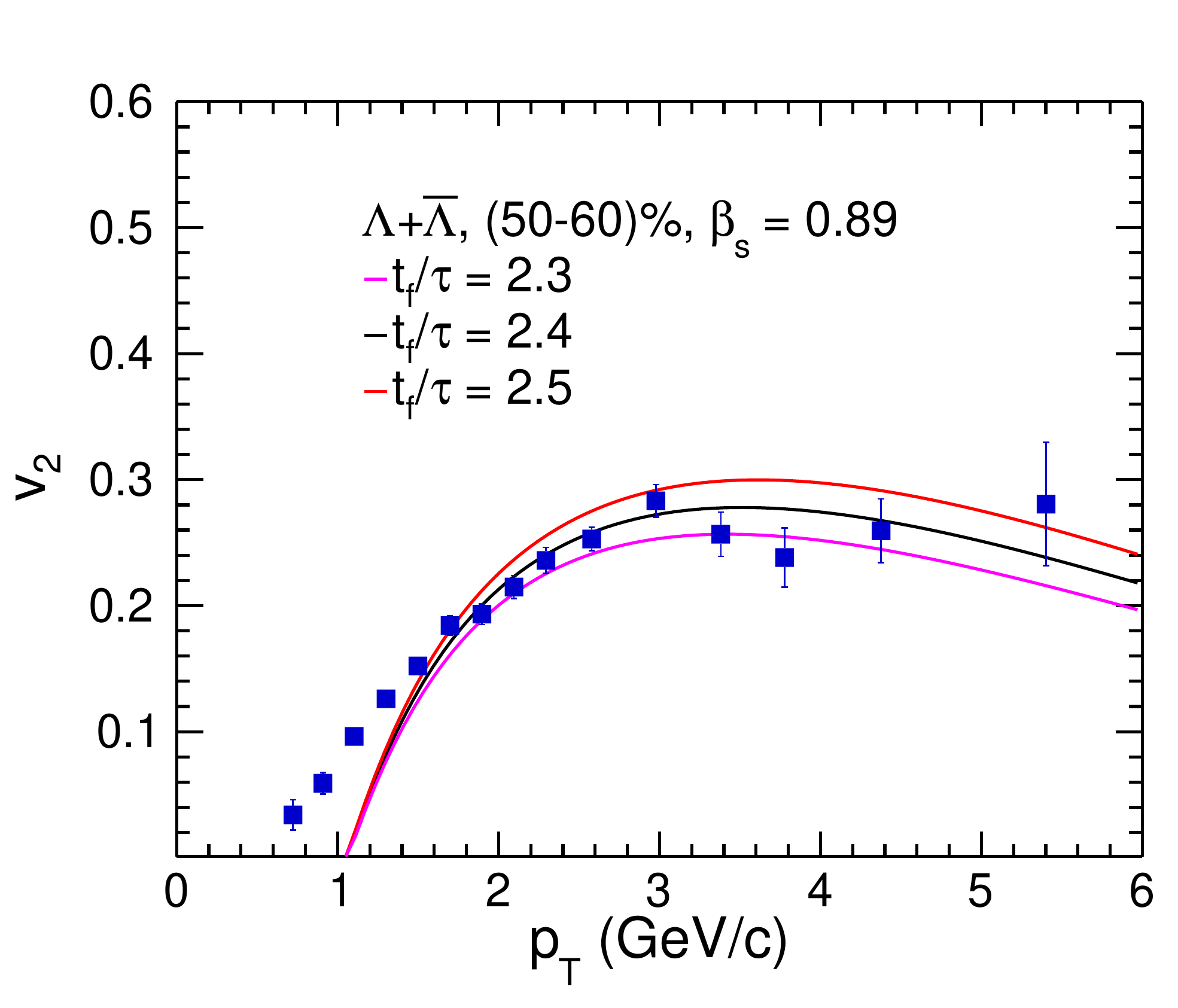}
\caption[]{(Color online) The elliptic flow ($v_2$) of ($\Lambda+\bar{\Lambda}$) versus $p_T$ at constant $\frac{t_f}{\tau}$ and $\beta_s$ for peripheral collisions (50-60)\% at $\sqrt{s_{NN}}$ = 2.76 TeV. Symbols are experimental data points~\cite{Abelev:2014pua} and lines are the model results.}
\label{lambdav2}
\end{figure*}

In fig.~\ref{lambdav2}, the elliptic flow of $\Lambda+\bar{\Lambda}$ is presented with respect to $p_T$ for centrality 50-60\% at $\sqrt{s_{NN}}$ = 2.76 TeV Pb+Pb collisions. In the left hand side of the figure, we show the $v_2$ for different values of $\beta_s$ keeping $t_f/\tau$ fixed. It is found that the model results explain the experimental data qualitatively above $p_T$ = 1 GeV/c for $\beta_s$ = 0.89. The right hand side of the figure is the results for various $t_f/\tau$ at constant $\beta_s$. Again there is a good agreement between the model calculations and experimental data above $p_T$ = 1 GeV/c for $t_f/\tau$ = 2.4.   

\begin{figure*}
\centering
\includegraphics[height=19em]{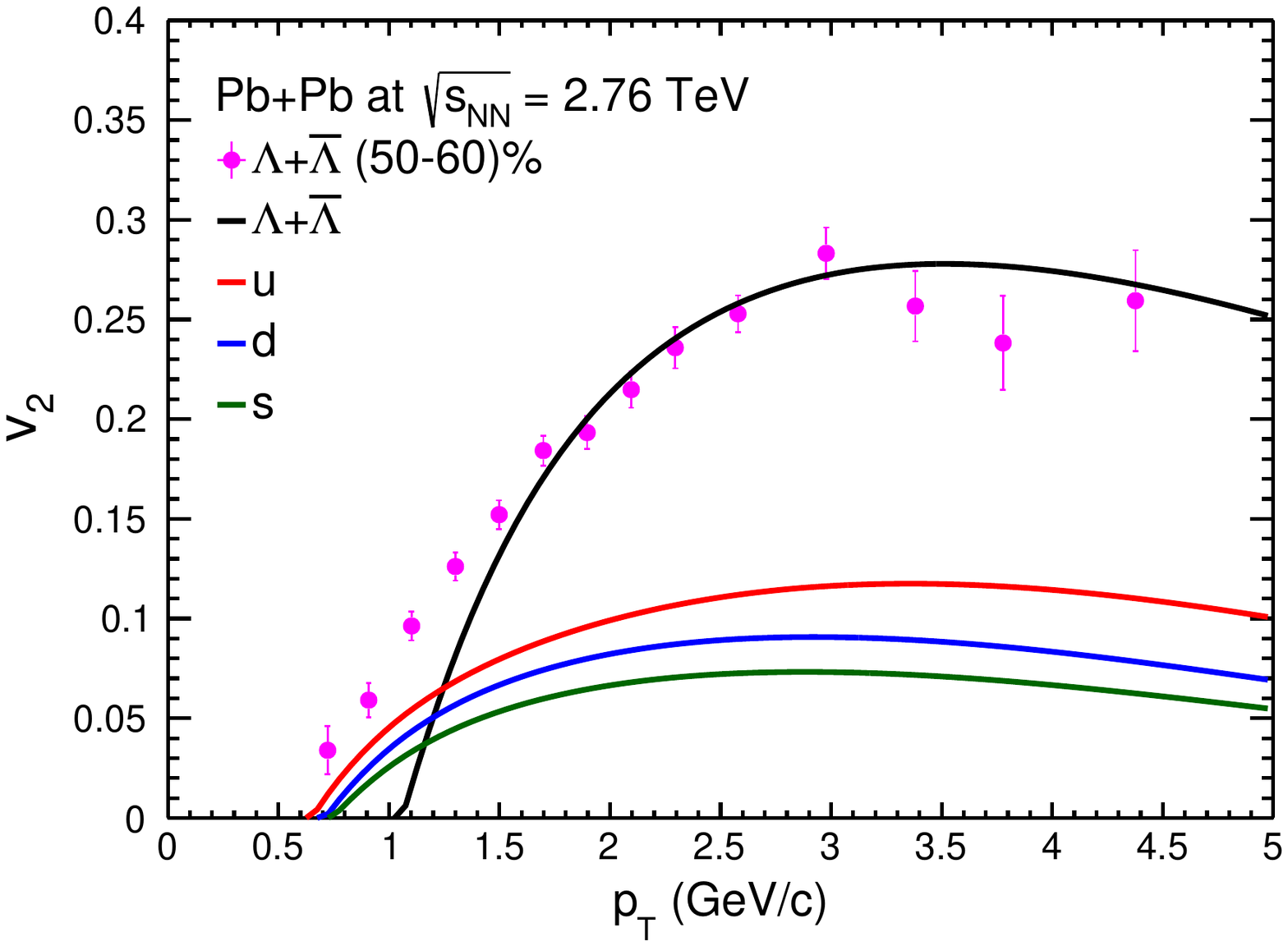}
\caption[]{(Color online) Comparison of $v_2$ of constituent quarks with the final $\Lambda$ hadron $v_2$ for peripheral Pb+Pb collisions at $\sqrt{s_{NN}}$ = 2.76 TeV.}
\label{lambdauds}
\end{figure*}

In fig~\ref{lambdauds}, we have plotted the $v_2$ of $\Lambda$ hadron with its three constituent quarks, $u,d,\text{and}\,s$. Although the flow of the constituent quarks start much before $p_T\,<$ 1.0 GeV/c unlike that of the $\Lambda$, the magnitude is much smaller than that of the hadron. Another which is visible from the plot is that the constituent quarks follow some sort of mass ordering with up quark being the lightest has highest flow and strange quark has the lowest. In this calculation $\beta_s$ and $t_f/\tau$ taken from $\Lambda$ $v_2$ plot are kept fixed for its constituent quarks, $u,d,\text{and}\,s$.

\begin{figure*}
\centering
\includegraphics[height=19em]{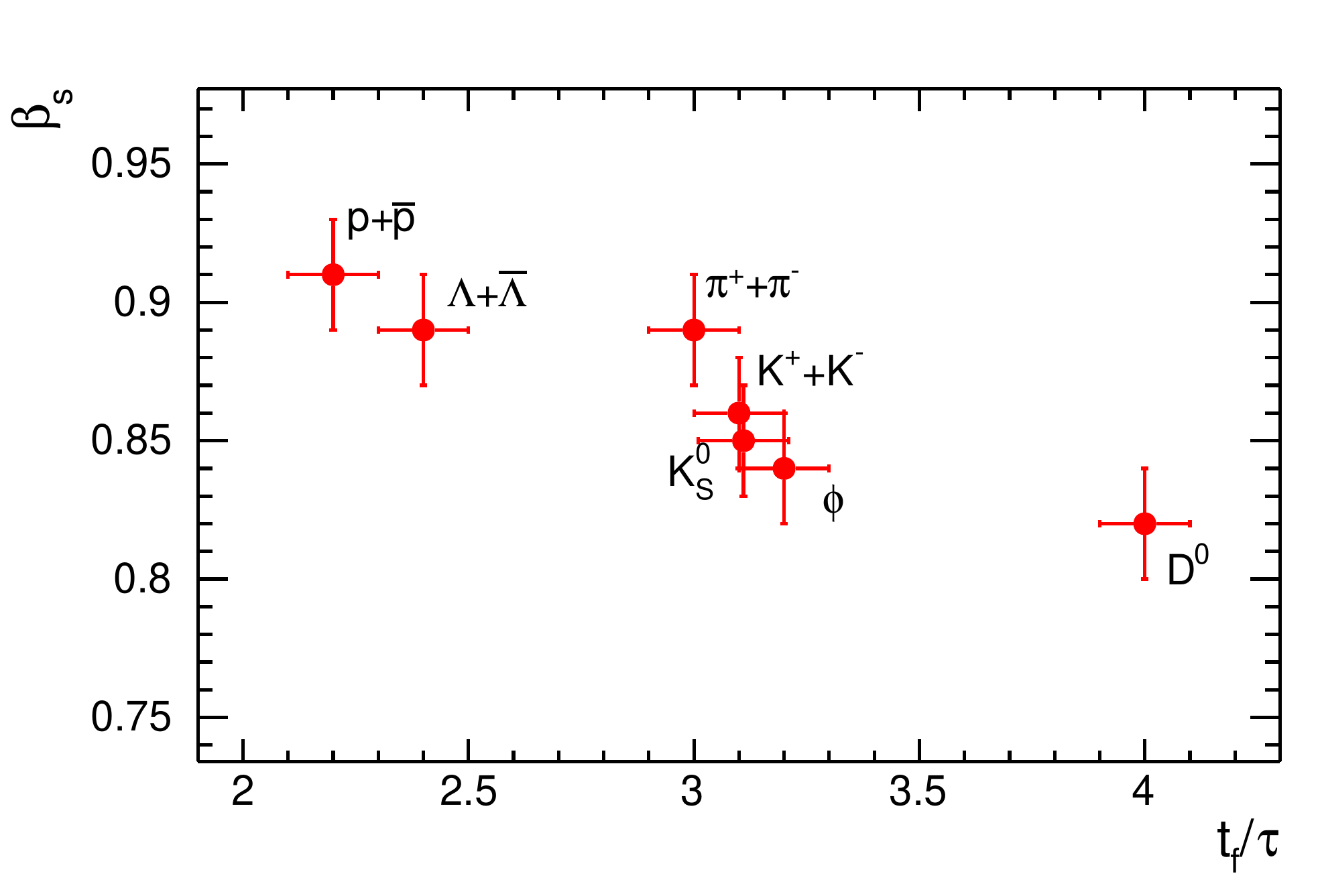}
\caption[]{(Color online) The plot of $\beta$ versus $\frac{t_f}{\tau}$ of various particles for peripheral Pb+Pb collisions at $\sqrt{s_{NN}}$ = 2.76 TeV.}
\label{betavstftau}
\end{figure*}

In fig.~\ref{betavstftau}, the correlation of $t_f/\tau$ with $\beta_s$ is shown for various identified hadrons observed after extracting the values from the model results on elliptic flow with the experimental data at $\sqrt{s_{NN}}$ = 2.76 TeV for peripheral Pb+Pb collisions. In this plot, we find that with the increase in $\displaystyle t_f/\tau$, the surface velocity, $\beta_s$ of hadrons decreases. The mesons show this trend separately from the baryons as evident from the figure. Although the ranges of variations in the values of both the parameters are not large, we find a small mass dependence in the correlation as we go from lightest $\pi$-meson towards heavier $D^0$ meson. Similar trend is also being observed for baryons, $p$ and $\Lambda$.

\section{Summary and Conclusions}
\label{Summary}
We have used quark coalescence method for hadronization and Boltzmann transport equation in relaxation time approximation to estimate elliptic flow, $v_2$ for the identified hadrons in Pb+Pb collisions at $\sqrt{s_{NN}}$ = 2.76 TeV. The important findings are summarised as follows:
\begin{enumerate}
\item  The quark coalescence approach is successful in explaining the elliptic flow data in the moderate transverse momentum region. However, it could not explain the data at low $p_T$.

\item The present formalism successfully attempts to connect the particle production from prompt interaction of initially produced partons with finally produced hadrons at hadronization hyper-surface. For intermediate p$_T$ ranges, the present formalism may successfully interpolate non-equilibrium or jet like quarks into blast wave distribution. The hadronic medium effects have not been taken into account. Similarly, resonance decays into observed particles particularly in pion sector have been neglected as well.  

\item We have found a correlation between the radial part of the transverse flow  and $\displaystyle t_f/\tau$ while explaining the $v_2$ spectra in peripheral collisions.

\item We have also compared elliptic flow of constituent quarks, ($u,d,s$) with the final hadron, $\Lambda$. We find that $v_2$ of each quark is around $\displaystyle 1/3$ of the final $\Lambda$ baryon. This actually verifies the coalescence mechanism used in the present calculations.   

\item Higher mass quarks are found to have a lower $v_2$ as compared to lighter quarks. On the other hand, flow of mesons behave almost similarly in the mid-$p_T$ region although their flow parameter, $\beta_s$ and time ratio, $t_f/\tau$ show correlation and a mass dependence. This is evident from the observation of monotonically decreasing flow parameter with time ratio and particle mass. This also shows that azimuthal anisotropy developed in the partonic phase plays a major role in the observed $v_2$ of final hadrons. Similarly, hadronization mechanism as a part of the freeze-out dynamics also play a major role in this regard. 

We will continue our investigation on particles' flow with other hadronization mechanisms such as fragmentation and compare with our current coalescence/recombination model, within the framework of BTE-RTA mechanism.

\end{enumerate}

\section*{Conflict of Interest}
The authors declare that there is no conflict of interest regarding the publication of this paper.

\section*{Acknowledgements}
The authors acknowledge the financial supports  from  ALICE  Project  No. SR/MF/PS-01/2014-IITI(G)  of  
Department  of  Science  \&  Technology,  Government of India. ST acknowledges the financial support by 
DST-INSPIRE program of Government of India.


\begin{thebibliography}{99}

\bibitem{voloshin}
S.~Voloshin and Y.~Zhang, Z. \ Phys. \ C {\bf 70}, 665 (1996).

\bibitem{LeFevre:2016vpp} 
  A.~Le Fèvre, Y.~Leifels, C.~Hartnack and J.~Aichelin,
  Phys.\ Rev.\ C {\bf 98}, 034901 (2018).
  
\bibitem{v2_star1}
J.~Adams {\it et al.} [STAR Collaboration],
  Phys.\ Rev.\ Lett.\  {\bf 92}, 052302 (2004).
  
\bibitem{v2_star2}
B.~I.~Abelev {\it et al.} [STAR Collaboration],
  Phys.\ Rev.\ C {\bf 75}, 054906 (2007).
  
\bibitem{v2_star3}
J.~Adams {\it et al.} [STAR Collaboration],
  Phys.\ Rev.\ C {\bf 72}, 014904 (2005).
  
\bibitem{v2_star4}
B.~I.~Abelev {\it et al.} [STAR Collaboration],
  Phys.\ Rev.\ Lett.\  {\bf 99}, 112301 (2007).
  
\bibitem{v2_phenix1}
S.~S.~Adler {\it et al.} [PHENIX Collaboration],
  Phys.\ Rev.\ Lett.\  {\bf 91}, 182301 (2003).
  
\bibitem{v2_phenix2}
S.~Afanasiev {\it et al.} [PHENIX Collaboration],
  Phys.\ Rev.\ Lett.\  {\bf 99}, 052301 (2007).
  
\bibitem{v2_ALICE1}
B.~B.~Abelev {\it et al.} [ALICE Collaboration],
  JHEP {\bf 1506}, 190 (2015).
    
\bibitem{Biro:2015iua} 
  T.~S.~Biró, M.~Horváth and Z.~Schram,
  Eur.\ Phys.\ J.\ A {\bf 51}, 75 (2015).  
  
\bibitem{Nahrgang:2014vza} 
  M.~Nahrgang, J.~Aichelin, S.~Bass, P.~B.~Gossiaux and K.~Werner,
  Phys.\ Rev.\ C {\bf 91}, 014904 (2015).
   
\bibitem{Haque:2017qhe} 
  M.~Rihan Haque, C.~Jena and B.~Mohanty,
  Adv.\ High Energy Phys.\  {\bf 2017}, 1248563 (2017).
  
\bibitem{Tripathy:2017nmo} 
  S.~Tripathy, S.~K.~Tiwari, M.~Younus and R.~Sahoo,
Eur.\ Phys.\ J.\ A {\bf 54}, 38 (2018).

\bibitem{Sun:2015pta} 
  X.~Sun, J.~Liu, A.~Schmah, S.~Shi, J.~Zhang, L.~Huo and H.~Jiang,
  J.\ Phys.\ G {\bf 42}, 115101 (2015).
  
\bibitem{Teaney:2000cw} 
  D.~Teaney, J.~Lauret and E.~V.~Shuryak,
  Phys.\ Rev.\ Lett.\  {\bf 86}, 4783 (2001). 
  
\bibitem{Mazumder:2013oaa} 
  S.~Mazumder, T.~Bhattacharyya and J.~e.~Alam,
  Phys.\ Rev.\ D {\bf 89}, 014002 (2014).
  
\bibitem{Sarkar:2017fni} 
  S.~Sarkar, P.~Mali and A.~Mukhopadhyay,
  Phys.\ Rev.\ C {\bf 96}, 024913 (2017).
  
\bibitem{Guo:2017mkf} 
  C.~Q.~Guo, C.~J.~Zhang and J.~Xu,
  Eur.\ Phys.\ J.\ A {\bf 53}, 233 (2017).
  
\bibitem{Sahoo:2018dxn} 
  P.~Sahoo, S.~K.~Tiwari and R.~Sahoo,
  Phys.\ Rev.\ D {\bf 98}, 054005 (2018)         

\bibitem{Baier:2006um} 
  R.~Baier, P.~Romatschke and U.~A.~Wiedemann,
  Phys.\ Rev.\ C {\bf 73}, 064903 (2006).
  
\bibitem{Gavin:1985ph} 
  S.~Gavin,
  Nucl.\ Phys.\ A {\bf 435}, 826 (1985).
  
\bibitem{Geiger:1991nj} 
  K.~Geiger and B.~Muller,
  Nucl.\ Phys.\ B {\bf 369}, 600 (1992).
  
\bibitem{Srivastava:1997qg} 
  D.~K.~Srivastava and K.~Geiger,
  Phys.\ Lett.\ B {\bf 422}, 39 (1998).
  
\bibitem{Bass:2004vh} 
  S.~A.~Bass, B.~Muller and D.~K.~Srivastava,
  J.\ Phys.\ G {\bf 30}, S1283 (2004).
  
\bibitem{Zhang:1999rs} 
  B.~Zhang, M.~Gyulassy and C.~M.~Ko,
  Phys.\ Lett.\ B {\bf 455}, 45 (1999).
  
\bibitem{Younus:2013rja} 
  M.~Younus, C.~E.~Coleman-Smith, S.~A.~Bass and D.~K.~Srivastava,
  Phys.\ Rev.\ C {\bf 91}, 024912 (2015).
      
\bibitem{Bhattacharyya:2015nwa} 
  T.~Bhattacharyya, P.~Garg, R.~Sahoo and P.~Samantray,
  Eur.\ Phys.\ J.\ A {\bf 52}, 283 (2016).

  
   \bibitem{Tripathy:2016hlg} 
  S.~Tripathy, T.~Bhattacharyya, P.~Garg, P.~Kumar, R.~Sahoo and J.~Cleymans,
  Eur.\ Phys.\ J.\ A {\bf 52}, 289 (2016).
  
  \bibitem{Tripathy:2017kwb} 
  S.~Tripathy, A.~Khuntia, S.~K.~Tiwari and R.~Sahoo,
  Eur.\ Phys.\ J.\ A {\bf 53}, 99 (2017).
  
  
\bibitem{Florkowski:2016qig} 
  W.~Florkowski and R.~Ryblewski,
  Phys.\ Rev.\ C {\bf 93}, 064903 (2016).
  
  \bibitem{Bjorken:1982qr} J.~D.~Bjorken, 
  Phys.\ Rev.\ D {\bf 27}, 140 (1983).
  
  \bibitem{Huovinen:2001cy} 
  P.~Huovinen, P.~F.~Kolb, U.~W.~Heinz, P.~V.~Ruuskanen and S.~A.~Voloshin,
  Phys.\ Lett.\ B {\bf 503}, 58 (2001).
  
  
\bibitem{Schnedermann:1993ws} 
  E.~Schnedermann, J.~Sollfrank and U.~W.~Heinz,
  Phys.\ Rev.\ C {\bf 48}, 2462 (1993).
  
  
\bibitem{BraunMunzinger:1994xr} 
  P.~Braun-Munzinger, J.~Stachel, J.~P.~Wessels and N.~Xu,
  Phys.\ Lett.\ B {\bf 344}, 43 (1995).
  
\bibitem{Tang:2011xq}
  Z.~Tang {\it et al.},
  Chin.\ Phys.\ Lett.\  {\bf 30}, 031201 (2013).
  
\bibitem{Adcox:2003nr} 
  K.~Adcox {\it et al.} [PHENIX Collaboration],
  Phys.\ Rev.\ C {\bf 69}, 024904 (2004).
  \bibitem{pythia} T.~Sjostrand, S.~Mrenna and P.~Z.~Skands,
  JHEP {\bf 0605}, 026 (2006).
   \bibitem{hijing} M.~Gyulassy and X.~N.~Wang, 
  Comput.\ Phys.\ Commun.\ {\bf 83} (1994) 307.
  
\bibitem{Lai:1996mg} 
  H.~L.~Lai, J.~Huston, S.~Kuhlmann, F.~I.~Olness, J.~F.~Owens, D.~E.~Soper, W.~K.~Tung and H.~Weerts,
  Phys.\ Rev.\ D {\bf 55}, 1280 (1997).
  
  \bibitem{Molnar:2003ff} 
  D.~Molnar and S.~A.~Voloshin,
  Phys.\ Rev.\ Lett.\  {\bf 91}, 092301 (2003).
  
  \bibitem{Lin:2003jy} 
  Z.~w.~Lin and D.~Molnar,
  Phys.\ Rev.\ C {\bf 68}, 044901 (2003).
  
  \bibitem{Fries:2003vb} 
  R.~J.~Fries, B.~Muller, C.~Nonaka and S.~A.~Bass,
  Phys.\ Rev.\ Lett.\  {\bf 90}, 202303 (2003).
  
  \bibitem{Fries:2003kq} 
  R.~J.~Fries, B.~Muller, C.~Nonaka and S.~A.~Bass,
  Phys.\ Rev.\ C {\bf 68}, 044902 (2003).

 \bibitem{Kaczmarek:1999mm} 
  O.~Kaczmarek, F.~Karsch, E.~Laermann and M.~Lutgemeier,
  Phys.\ Rev.\ D {\bf 62}, 034021 (2000).
  
  \bibitem{Peshier:2002ww} 
  A.~Peshier, B.~Kampfer and G.~Soff,
  Phys.\ Rev.\ D {\bf 66}, 094003 (2002).
  
  \bibitem{Casalderrey:2003cf} 
  S.~J.~Casalderrey and E.~V.~Shuryak,
  hep-ph/0305160.  
  
    \bibitem{Greco:2003mm} 
  V.~Greco, C.~M.~Ko and P.~Levai,
  Phys.\ Rev.\ C {\bf 68}, 034904 (2003).
  
  \bibitem{Sun:2017ooe} 
  K.~J.~Sun and L.~W.~Chen,
  Phys.\ Rev.\ C {\bf 95}, 044905 (2017).
  
  \bibitem{He:2017tla} 
  Y.~He and Z.~W.~Lin,
  Phys.\ Rev.\ C {\bf 96}, 014910 (2017).
  
  \bibitem{Yin:2017qhg} 
  X.~Yin, C.~M.~Ko, Y.~Sun and L.~Zhu,
  Phys.\ Rev.\ C {\bf 95}, 054913 (2017).
  
  \bibitem{Peitzmann:2005ty} 
  T.~Peitzmann,
  Acta Phys.\ Hung.\ A {\bf 27}, 363 (2006).
  
  \bibitem{GKuipers:MasterThesis}
  G.~Kuipers, Advisors: E.~Laenen and T.~Peitzmann, (2005). 
  \href{https://www.nikhef.nl/pub/theory/masters-theses/gerben\_kuipers.pdf}{https://www.nikhef.nl/pub/theory/masters-theses/gerben\_kuipers.pdf}
 
\bibitem{Abelev:2013vea} 
  B.~Abelev {\it et al.} [ALICE Collaboration],
  Phys.\ Rev.\ C {\bf 88}, 044910 (2013).
  
  \bibitem{Greco:2004ex} 
  V.~Greco and C.~M.~Ko,
  Phys.\ Rev.\ C {\bf 70}, 024901 (2004).

  \bibitem{Abelev:2013xaa} 
  B.~B.~Abelev {\it et al.} [ALICE Collaboration],
  Phys.\ Rev.\ Lett.\  {\bf 111}, 222301 (2013).
  
  \bibitem{Abelev:2014laa} 
  B.~B.~Abelev {\it et al.} [ALICE Collaboration],
  Phys.\ Lett.\ B {\bf 736}, 196 (2014).
   
\bibitem{Adam:2017zbf} 
  J.~Adam {\it et al.} [ALICE Collaboration],
  Phys.\ Rev.\ C {\bf 95}, 064606 (2017).

\bibitem{Nonaka:2003ew} 
  C.~Nonaka, B.~Muller, M.~Asakawa, S.~A.~Bass and R.~J.~Fries,
  Phys.\ Rev.\ C {\bf 69}, 031902 (2004).
  
\bibitem{He:2010vw} 
  M.~He, R.~J.~Fries and R.~Rapp,
  Phys.\ Rev.\ C {\bf 82}, 034907 (2010).

\bibitem{Abelev:2014pua} 
  B.~B.~Abelev {\it et al.} [ALICE Collaboration],
  JHEP {\bf 1506}, 190 (2015).

\bibitem{Adam:2015sza} 
  J.~Adam {\it et al.} [ALICE Collaboration],
  JHEP {\bf 1603}, 081 (2016).
  
  \bibitem{Abelev:2013lca} 
  B.~Abelev {\it et al.} [ALICE Collaboration],
  Phys.\ Rev.\ Lett.\  {\bf 111}, 102301 (2013).
  
  \bibitem{Miller:2007ri} 
  M.~L.~Miller, K.~Reygers, S.~J.~Sanders and P.~Steinberg,
  Ann.\ Rev.\ Nucl.\ Part.\ Sci.\  {\bf 57}, 205 (2007).
  

\end{thebibliography}
\end{document}